\documentstyle[12pt]{article}
\begin{document}
\pagestyle{plain}

\title{\bf CREATIVITY LEADING TO DISCOVERIES IN PARTICLE  PHYSICS AND 
RELATIVITY}

\author{\bf Miroslav Pardy\\[3mm]
Department of physical electronics \\
and\\
Laboratory of plasma physics\\
Faculty of Science, Masaryk University\\
Kotl\'{a}\v{r}sk\'{a} 2, 611 37 Brno, Czech Republic}
\date{\today}
\maketitle

\begin{center}
{\bf Abstract}
\end{center}

\vspace{1cm}
\baselineskip 14pt

Independently on Popper's, Holten's and Kuhn's philosophy of science,
we present -- in ``annus mirabiliss 2005'' --
the basic ingredients of discovery creativity in physics. We discuss 
understanding,
problem solving,  heuristics, computer thinking, technological
thinking and so on. We present some discoveries from the viewpoint of
creativity. The Dirac equation, the Riccati equation for
massive photons in laser physics, the nonlinear Schr\"{o}dinger
equation and its classical limit for heavy particles, the quantum
Navier-Stokes equation, the equation of the quantum
magneto-hydrodynamics and so on.  We discuss general relativity,
the nonlinear Lorentz transformation involving the maximal
acceleration constant which we relate to the Hagedorn temperature,
and possible dependence of mass on acceleration. We discuss the
reciprocity of technology and theoretical physics and the
technological limit.

\hspace{3ex}
\section{Introduction}

Michael Faraday -- man of simplicity --
 used special experimental methods to discovery many
laws of nature including famous magnetic induction. Thomas Alva Edison used
specific experimental techniques and method "trial and error" to
generate inventions useful for
daily life. Nicola Tesla invented electrical motor for alternate
current. Euler, Poincar\'e,  and others realized
mathematical discoveries every day.
All these thinkers  used undoubtedly some heuristical rules, or ``road map''
to achieve success.

What are the  dominating factors which determine the
discoveries and the key discoveries? What are the specific steps?
What are the tenacious ideas, or, purposes  which are leading  to
the key discoveries?

The principles of discoveries
are not always accepted by physicists. Why? Because we are
in a jungle of facts and the only way forwards we built is the way by
{\bf trial and error}.

On the other hand, if we want to avoid disappearing of the
total or partial content of the classical physics in the new
theory it is desirable {\bf  only small change of the classical
theory and to built the new theory as a paradigm of the old theory}.
Such a principle was used by Bohr in his planetary model of
atom. However, de Broglie and Schr\"{o}dinger deflected from the
Bohr instructions. Similarly, in particle physics, quark theory  --
nuclear aristocracy -- is not small
change of the bootstrap theory -- nuclear democracy --, but revolution.

The integral part of the process  of discoveries is, no doubt,
also an effort of the nameless and godforsaken investigators who prepared
the way to a discovery. In
the scope of this process a discoverer appears as a single man among a large
team of scholars who has completed the initiated work.

We know from history (Hadamard, 1946) that the
predecessor of the infinitesimal
calculus was Nicole Oresme (1323-1382), Johann Kepler (1571-1630) and
Pierre Fermat (1601-1665), the author of well-known famous theorem.
Fermat received the mathematical expression for
the condition of maximum and minimum of a function. He constructed expression

$$\frac {f(a + \varepsilon) - f(a)}{\varepsilon},
\quad \varepsilon \rightarrow \quad 0. \eqno(1)$$

However, this expression is nothing else
than the derivative of $f(x)$ at point $x =  a$, and Fermat also applied it
to finding tangents to several curves. However, nor Fermat was able to
overcome an unknown barrier standing at the way to the infinitesimal
calculus (Hadamard, 1946). Oresme, Kepler and Fermat realized
only the first step of the creation of the infinitesimal calculus.

It is necessary to distinguish two steps (Hadamard, 1946).

\begin{quote}
{\bf ``To discover a fact on the one hand, and to understand the
significance of this fact on the other hand''}.
\end{quote}

When Newton and Leibniz appeared in the
sky of mathematics, the infinitesimal calculus was realized.
While Fermat applied his method for special examples,
Newton and Leibniz gave a precise form to that idea, in driving it far
enough to be able to make it a starting point for further researches.

The decisive
factors of the birth of the infinitesimal calculus were the definition of
the notion of derivative, which is of course not sufficient,
and the application
of this mathematical object to the large amount of problems.
Thus, the discovery of infinitesimal calculus was based, as we can deduce,
on the two general principles. {\bf Introducing the new notion and its
large application.}

The misunderstanding of the significance of the fact is well know in
particle physics where the so called $\Omega^{-}$-meson  was
predicted in 1962 as the crucial test of the $SU(3)$ symmetry and
discovered in Brookhaven in 1964. However, already in 1954 Eisenberg
had observed $\Omega^{-}$-meson and in 1955 it was confirmed by
Fry. The laboratory events was interpreted as the $\Omega^{-}$ events
and not as the  $\Omega^{-}$-meson. The experimenters were not
prepared to understand the significance of the discovery of the
new particle (Alvarez, 1973). Similar events in
the particle physics related to the Popper philosophy of discoveries
was  presented  in the Pietschmann (1978) article.

At present time the excited states of proton is the brilliant
confirmation of the quark composition of proton. But the society
consciousness hesitates in  understanding of this fact.
The excited states of an electron is
the manifestation of its substructure but the society
consciousness hesitates in understanding of the significance of this fact.

In physics, Bohr created a model of atom bearing
later his name. Bohr created two postulates
which define the model of atom: 1. every atom can exist in the discrete
series of states in which electrons do not radiate even if they are moving at
acceleration  (the postulate of the stationary states), 2. transiting electron
from the one stationary state to other emits energy  according to
the law

$$\hbar\omega = E_{m} - E_{n},\eqno(2)$$
where $E_{m}$ is the energy of an electron in the
initial state, and $E_{n}$ is the energy of the final state of an
electron  to which the transition is made and $E_{m} > E_{n}$.
This postulate involves also the assumption that the photons are created by
transitions and they exists only outside of atom. Photons are real
particles, real quanta as was confirmed by the Compton effect.  {\bf There
is no photon inside the atom.} The multiphoton Compton effect is also
possible as follows from the Feynman diagrams, or from the Volkov solution of
the Dirac equation (Berestetzkii, 1989; Pardy, 2003a, 2003d, 2004b).
This effect was verified by high energy laser experiments.

The photon situation is reciprocal to the quark situation in particle physics.
{\bf Quarks exist only inside the mesons and baryons. New  mesons, or
  baryons are created and cloned inside mesons and baryons by the
tension of the physical strings connecting quarks.}
The existence of quarks inside mesons and hadrons was verified by
many deep inelastic scattering experiments.

The Planck distribution law of photons inside the so called black body was
derived by Einstein from the Bohr model. Einstein
introduced coefficients of spontaneous  and stimulated emission
$A_{mn}, B_{mn}, B_{nm}$. From the condition of equilibrium

$$N_{m}A_{mn} + N_{mn}\varrho_{\omega}B_{mn} =
N_{n}\varrho_{\omega}B_{nm} \eqno(3)$$
and the Maxwell statistics

$$N_{m} = De^{-\frac{E_{m}}{kT}},\quad  N_{n} = De^{-\frac{E_{n}}{kT}},
\eqno(4)$$
he derived the Planck law in the form:

$$\varrho_{\omega} = \frac{\hbar\omega^3}{\pi^{2} c^3}\frac{1}
{e^{\frac{\hbar\omega}{kT}} - 1}. \eqno(5)$$
The decisive point of the derivation of the black body by Einstein
was not the mathematical beauty of the equations but specific model
suggested by intuition.

However, the Einstein derivation was not complete, because he did not
consider the situation that the black body can be  influenced by some
external field. Then, it is necessary to introduce coefficients of the
external influence $ C_{mn}, C_{nm},$ and from the equilibrium
condition to derive the generalized Planck law (Pardy, 2005c).

$$ \varrho_{\omega} = \frac{\hbar\omega^3}{\pi^{2} c^3}\frac{1}
{e^{\frac{\hbar\omega}{kT}} - 1} +
\frac{\hbar\omega^3}{\pi^{2} c^3}\;\frac{P(\omega) -
 Q(\omega)e^{\frac{\hbar\omega}{kT}}}{e^{\frac{\hbar\omega}{kT}} - 1}
\eqno(6) $$
where $P, Q$ are some function which must be calculated using the
advanced solid state physics and advanced quantum mechanics.

In the special theory of relativity, Einstein was not influenced by the
partial results of his predecessors. Namely by,
W. Voight (1850-1919) who proposed in the year 1887 an
idea of the transformation of coordinates and time, J. Larmor who in
1900 derived independently so called Lorentz transformation and H.
Poincar\'e (1854-1912) who derived  in 1905 the similar consequences
as Einstein.

Einstein worked independently. The dominating
factors  leading Einstein to his theory were  admiration of the
axiomatic  system of Euclidean geometry, with the goal to unify the
electromagnetism and moving bodies by the
axiomatic way,  and a courage to postulate a principle of constant
light velocity as the integral part of his axiomatic system. This
principle is from  the viewpoint of common sense, very strange and
nobody was of a courage to postulate such strange
principle. {\bf Einstein decided to hit the nail on the head}.

The reasonable generalization of the special theory of
relativity for noninertial
systems and gravity, respectively, was a theory working with the
non-Euclidean mathematical objects. Einstein decided to apply
the Riemann geometry
which was yet developed by B. Riemann (1826-1866) and T. Levi-Civita (1873-1941).

Dominating heuristic factors in the birth of the
general relativity theory were the Einstein goal to unify STR and
gravity. Later, Hilbert proved that Einstein gravity equations could  be
easily derived from variation principle with the adequate Lagrangian.
Why was not so
easy to create general relativity?  Leibniz, Kant,
Schopenhauer and others created the statements that space and time
is only the subjective feelings, or, \'{a} priori auxiliary notions,
and not objective reality.
Loba\v{c}evskii and Gauss introduced  space and time as  physical
reality. This was the starting point of the modern science of space and time
and of the Einstein general relativity.

Now, let us look more to creativity in theoretical physics
leading to discoveries.

\section{On creativity in general}

The theoretical physics was realized by the creative work
and not by copying the older physical texts and existing products.
The creative work is a novel work.
It can be meaningful or meaningless. It is meaningful in
the human sense, if it is accepted by the society, or
by the specific group of people as useful and valuable work. The
"novel work" means that the creative product did not exist previously
in the same form. It arises from the already existing material by the
process of reorganization, reintegration, modification, new
axiomatization, redefinition of basic parts and so on,  in such a way that when
it is completed it contains elements that are new. The novelty of the
creative work is given by the novelty of methods
and ideas used and by the
deviation from the traditional knowledge or the status quo.
The novelty of the creative work depends usually on the novelty of the
problem and
especially on the creative individual. The creative individual
has the specific motivation to create. The creativity motivation does
not differ from any motivation problem. In other words, the individual
experiences a state of disequilibrium, or a lack of satisfaction with
the existing state of affairs. In such a way the homeostasis of the
creative individual is disturbed. This is possible only when there
exists the sensitivity to the given state. The creative person has a
lower threshold, or the greater sensitivity for the gaps or the lack
of closure that exist in the environment (Bodnarczuk, 1990).
The sensitivity is
obviously given by the intellectual factors, involving fantasy,
inborn in the creative person
and by the educational drill which cultivate the mental life of a person.
It also means that computer creativity can be only mechanical one and
never will involve fantasy, because fantasy cannot be realized by
non-stochastic computer algorithm.

Mathematics -- art for art -- is the
science of the pure forms which gives no propositions about outer
world. Theoretical physics, is, as can be seen later, the applied
pure mathematics. It means that theoretical physics is pure mathematics
restricted by the physical experiment and involving physical interpretation.

While mathematician starts from pure definitions and then proceeds to
show how the product of thought may originate either by
single act of creation or by the double act of positing and
combining, the theoretical physicist interprets mathematical
definitions (for instance operators in quantum mechanics, Green's
functions in the quantum field theory, and so on) and expresses the
outer world in the mathematical form.

However, because the theoretical physicist works practically with the
mathematical objects, his thinking is the mathematical one, but
limited by the outer world laws. Nevertheless he uses the same
heuristics as mathematician. Then, the principles of creativity in
mathematics and theoretical physics are practically congruent.
The effectiveness of work in physics and mathematics
consists in clarification and using these principles.

The interesting is the opinion of Julian Schwinger on the creativity.

\begin{quote}
{\bf " What the physicist actually does? He does not begin with the
final theory and draw mathematical consequences, he elaborates the
theory, he describes a certain class of phenomena and then
extrapolates and sees how far it works"}.
\end{quote}

Every rigorous scientific work is based
on the operative rules which are rigorously defined or
are only intuitive ones, for which man failed to obtain a precise algorithm.
If we use the intuitive rules in
science we say that we use the heuristic procedures. The aim of the
heuristic procedures, or the heuristics is to solve new problems
for which the algorithmic solutions are not known, or difficult to
find, and to make discoveries.

\section{On theoretical methods in physics}

One can distinguish between two main procedures in theoretical physics.
The first one is to work with experimental facts which means
the close contact with the experimental physicists.
This method involves the systematical collection of the experimental
data and trial to fit them into a comprehensive and satisfying scheme.

The other procedure is to work from the mathematical basis.
This second method was in the modern physics used in the maximally
effective way by P. A. M. Dirac. In this method one examines and
criticises the existing theory. One tries to pin-point the faults in it and
then tries to remove them by replacing it by the new scheme.
The obstacles in the second way consist in removing the faults
without destroying the great successes of the existing theory.
In other words,  the second procedure is the replacing of the one theory
by the better one in such a way that the better theory involves some
details of the first theory, other details distracts and implants
into the new schema some new ideas.

While the first method called phenomenology requires the common sense and not
very much mathematics, the second method is only for the mathematical
experts who use the mathematics with the goal to understand nature.

The procedure which works from the mathematical
basis was sometime also unsuccessful. For instance, the gravitational and the
electromagnetic fields should be closely connected, but Einstein spent
many years trying to unify them, without success. Nevertheless
it is necessary to say that the Einstein endeavor formed the basis
for the unification of electromagnetic and weak interactions and for
the struggle to unify the all forces in nature into the grand
unification scheme.

Whether one follows the phenomenological or the mathematical procedures
depends on the subject of study and on the intellectual mentality
of the investigator. Let us illustrated it by the discovery
of quantum mechanics (Salam).

While Heisenberg was working from the experimental basis, using the
results of spectroscopy which by 1925 had accumulated an enormous
amount of experimental data, Schr\"{o}dinger worked from the mathematical
basis. Among these data were some very important, namely the relative
intensities of the lines of the multiplets. Only Heisenberg was able to
pick up the substantial thing from the great wealth of data and arrange them
in a matrix scheme, which led to the matrix quantum mechanics.

On the other hand Schr\"{o}dinger had the idea at the back of his mind that
spectral frequencies should be fixed by eigenvalue equation, something like
those that fix the frequencies of system of vibrating strings.

Of course, the Schr\"{o}dinger spectral idea was an integral part
of his intellectual attitude to modern physics especially to the
relativity. Schr\"{o}dinger was influenced by the new physical ideas
and so he tried to set up a quantum mechanics within the framework of
relativity. While de Broglie applied the wave idea to the free particles,
Schr\"{o}dinger tried to generalize it to an electron bound in atom within the
relativistic framework. However, when he applied his theory to the
hydrogen atom, he found it did not agree with experiment. The
discrepancy was caused by the ignoring the spin of the electron.
Then Schr\"{o}dinger noticed that his theory was correct in the
nonrelativistic approximation which forms the well known Schr\"{o}dinger
equation.

\section{The Dirac heuristics}.

Nobel Laureate Abdus Salam says on Dirac (Kamran, 1989):

\begin{quote}
{\bf "Paul Adrien Maurice
Dirac was undoubtedly one of the greatest
physicists of this or any century. In three decisive years,
1925, 1926  and 1927, with three papers, he laid the foundations first
of quantum physics, second of the quantum theory of fields, and
third of the theory of elementary particles with his famous equation
of the electron. No man except Einstein has such a decisive
influence in so short a time on the course of physics in this century"}.
\end{quote}

So, if Dirac is compared with Einstein, then is is worth while to know
his heuristic methods in theoretical physics
used by him to perform such big discoveries.
Many Dirac authentic statements concerning
the methods of discoveries in theoretical physics are very useful for
understanding his methods of discoveries. Surprising is that {\bf Dirac
excluded the philosophical thinking in his theoretical works}
(Kamran, 1989)).

\begin{quote}
{\bf "I then felt that all
the things that philosophers said were rather indefinite, and I came to the
conclusion eventually that I did not think philosophy could
contribute anything to the advance of physics".}
\end{quote}

The basic philosophical question is what does a word mean or what does
every phrase in the verbal expression of thought mean? Just such
attitude but in the more mathematical form is involved in the Dirac
modes of thought.

The very instructive are the Dirac opinions on the Einstein general theory
of relativity and on the magnetic monopole (Kamran, 1989).

\begin{quote}
{\bf "One has a great confidence in
the theory arising from its great beauty, quite independent of its
detailed success".}
\end{quote}

\begin{quote}
{\bf  ".. monopoles should exist, because of
pretty mathematics".}
\end{quote}

The pretty mathematics is also the integral part of Dirac's derivation of his
famous equation. The heuristic procedure which Dirac used was so
called {\bf playing with equation}. The discovery of the Dirac equation was
realized after the Pauli equation which Pauli proposed in May 1927
(Pais, 1986).

The Pauli equation describes the electron spin by the two-component spinors
in such a way that spin is explicitly coupled to the electron angular
momentum. The strength of the coupling called the "Thomas factor" cannot be
determined inside the Pauli theory and it must be inserted by hand
without further justification. Pauli noted that such unpleasant feature
is caused by its nonrelativistic behavior. In other words Pauli was award
that this equation was approximate and provisional.

The spin was described in the Pauli theory by the  2 x 2  matrices, since known
as the the Pauli matrices. However, now it is known that Dirac discovered
these independently. Dirac writes (Dirac, 1977b):

\begin{quote}
{\bf "I believe I got these matrices
independently of Pauli, and possibly Pauli also got them independently
from me".}
\end{quote}

However Dirac used matrices of four rows and columns.
At this procedure Dirac discovered that his  4 x 4  matrices were
composed from the Pauli  2 x 2  matrices.

Thus the Dirac equation was born in 1928 (Dirac, 1928a, 1928b)
by respecting the special
theory of relativity and using the heuristics {\bf linearization of the
nonlinear operator}. While the rigorous relativistic generalization of
the Schr\"odinger equation is

$$i\hbar\frac{\partial\psi}{\partial t} = \sqrt{c^{2}{\bf p}^{2} +
m^{2}c^{4}}\; \psi. \eqno(7)$$
Dirac use the matrix transcription where it is possible  to write

$$\sqrt{c^{2}{\bf p}^{2} + m^{2}c^{4}} = 
c\sum_{\mu = 1}^{4}\alpha_{\mu}p_{\mu},\eqno(8)$$
where $\alpha_{\mu}$ are composed from Pauli
matrices. Dirac did not introduce numbers $p, q, r, s$ in such a way
that $(a^2 + b^2 + c^2 + d^2) = (pa + qb + rc + sd)^2$ and did not
define special relations between numbers  $p, q, r, s$. He used
matrices, because Pauli equation involved matrices and successfully
described the quantum nonrelativistic motion an electron. In case of
introduction of numbers $p, q, r, s$, we get the relativistic
Schr\"{o}dinger equation of the first order while  Klein-Gordon equation
is equation of the second order.

The great Dirac discovery he has made was, that his equation gave the particle
a spin of half a quantum. Also gave it a magnetic moment. It gave
just the properties of an electron (Dirac, 1977b).

The obvious consequences of the Dirac equation were the Sommerfeld fine
structure formula and the Thomas factor. For energies small compared
to $m\* c^2$ all the results of the nonrelativistic Schr\"{o}dinger
theory were recovered.

While at this time the Dirac equation is famous, the serious objections
were to this equation  at time of its birth. The fundamental objection was
the existence of the negative energies of the electrons which followed
inevitably from it. Now, they are interpreted as the positron energies,
where positron is the antiparticle to electron. Dirac himself supposed
that his equation describes proton instead of positron, at this time
nondiscovered particle. However, the magnetic moment of proton differs
from the Dirac prediction.

Dirac was never satisfied quite with his equation and in year
1972 postulated the new equation with the positive energy
solution only (Dirac, 1972a, 1972b, 1977a).
Although the new equation is also studied at the present
time, it is not so famous as the original equation from year 1928.

So we have seen that the specific principles guiding Dirac to his
discovery were the pretty mathematics, playing with equation and the
principles of the special theory of relativity. During the discussion with
Abraham Pais on the research methods, Dirac said (Pais, 1986):

\begin{quote}
{\bf  "First play with
pretty mathematics for its own sake, then see whether this leads to
new physics".}
\end{quote}

The method of "playing with mathematics" was the "leitmotiv" of
the all Dirac life around. At age 28 he writes (Dirac, 1931):

\begin{quote}
{\bf " The most powerful method of advance
that can be suggested at present is to employ all the resources of pure
mathematics in attempts to perfect and generalize the mathematical
formalism that forms the existing basis of theoretical physics, and
after each success in this direction, to try to interpret the new
mathematical features in terms of physical entities".}
\end{quote}

At age 36 Dirac writes (Dirac, 1939):

\begin{quote}
{\bf "As time goes on it becomes increasingly evident
that the rules which the mathematician finds interesting are the same
as those which Nature has chosen".}
\end{quote}

At age 60 he writes (Kuhn, 1963):

\begin{quote}
{\bf "I think it's a peculiarity of myself that I like
to play with equations, just looking for beautiful mathematical relations
which may be don't have any physical meaning at all. Sometimes they do".}
\end{quote}
At age 78 he writes (Dirac, 1982):

\begin{quote}
{\bf "A good deal of my research work in physics has
consisted in not setting out to solve some particular problems, but simply
examining mathematical quantities of a kind that physicists use and trying
to fit them together in an interesting way regardless of application that
the work may have. It is simply a research for pretty mathematics. It may
turn out later that the work does have an application. Then one has
good luck".}
\end{quote}

So we have seen that Dirac was deeply convinced that the best method how
to find truth in Nature is the pretty mathematics.
So, mathematics is for the physicists not only the
instrument of the description of the phenomena,
however, it is the main source of ideas and principles of the new
theory. {\bf With the help of the pure mathematical
construction we can find the notions and the relations between them
which give us the key in order to understand the phenomena in nature.}
The experiment remains the only single criterion of the adequacy, or
non-adequacy of the mathematical constructions. However, the present
creative basis belongs to mathematics.

The foregoing statements do not mean that if one is expert in mathematics,
he is at the same time theoretical physicist because there is the
substantial difference between mathematics and theoretical physics. While the
mathematical statements are according to Bertrand Russell - the
greatist philosopher of england -
non-existential, the physical statements inform
us on the existing relations and things in nature.

The mathematical program can be effective in the
theoretical physics if it involves the deep physical principles and ideas.
From this point of view it is possible to analyze the effectiveness
of such theories as the  Regge theory of elementary particles, the bootstrap
theory, the supersymmetry, supergravity, source theory, the present
time string theory and so on. Only if they are based on the deep
physical ideas, then,  there is some chance to be physically true.

Let us summarize the discourse on the Dirac heuristics by his own
statement (Dirac, 1978).

\begin{quote}
{\bf "I learn to distract all physical concepts as the basis for
theory. Instead one should put one's trust in a mathematical scheme,
even if the scheme does not appear at first sight to be connected with
physics. The physical meaning had to follow behind the mathematics".}
\end{quote}

Now, we can show that the Dirac method enables also the playing with his
equation itself (Pardy, 1973a). Let us write his equation in the form:

$$\varphi_{1} + \varphi_{2} + mc\psi = 0,
\eqno(9)$$
where

$$\varphi_{1} = \hbar\gamma_{\mu}\frac{\partial\psi}{\partial x_{\mu}};
\quad  \varphi_{2} = -i\frac{e}{c}A_{\mu}\gamma_{\mu}\psi. \eqno(10)$$

We can multiply the last equation from the left by gamma-matrices  $\Gamma$
an by the conjugated function $\bar\psi = \psi^{+}\gamma_{4} =
(\psi_{1}^{*},\psi_{2}^{*},\psi_{3}^{*}, \psi_{4}^{*},)$ in order to
get:

$$\bar\psi\Gamma_{\alpha}
(\varphi_{1} + \varphi_{2} + mc\psi) = 0,\eqno(11)$$
where

$$\gamma_{\alpha} = I, \gamma_{\mu}, \sigma_{\mu\nu},
i\gamma_{5}\gamma_{\mu}, \gamma_{5}\eqno(12)$$
with

$$\sigma_{\mu\nu} = \frac{1}{2i}(\gamma_{\mu}\gamma_{\nu} -
\gamma_{\nu}\gamma_{\mu}).\eqno(13)$$

It is possible to show that the equation (11) is equivalent to the
Dirac equation (Pardy, 1973a). Its mathematical beauty corresponds to
the beauty of the Dirac equation.

The meaning of the above procedure is, that electron can be evidently
described by the components $\bar\psi\Gamma_{\alpha}\psi$.
It is possible to derive the Lorentz
equation for the classical motion of a point particle in the
electromagnetic field for the component $\bar\psi\gamma_{\mu}\psi$.
The so called Bargmann-Michel-Telegdi spin equation
follows for the component
$\bar\psi\gamma_{\mu}\gamma_{5}\psi$ (Rafanelli et al. 1964; Pardy, 1973a).
The procedure performed with Dirac equation can be applied in the
analogical form to the Duffin-Kemmer equation (Pardy, 1973b).

There is a case when Dirac equation generates new equation. We think
the case of the Volkov solution of the Dirac
equation for an electron moving in the  electromagnetic field
of massive photons. While in case for the massless photons the
solution is known,  the massive photon situation leads to the so
called Riccati equation (Pardy, 2003d, 2004d)

$$y' + y^2 + P(x) = 0, \eqno(14)$$
where the function $y$ is the integral part of the spinor $\psi$. So,
the Dirac equation generates the new differential equation which, as
it is known, has no solution in the closed form in general.
The Riccati equation
was mathematically meaningful from the time of its discovery. Now,
it is meaningful also from the view point of physics (Pardy, 2004d).

It can be easily to see that the playing with equation and with mathematics
of the nonrelativistic quantum mechanics leads to the new generalized
quantum mechanical equation of the nonrelativistic quantum mechanics.
Let us demonstrate it here.

If we add to the original equations of hydrodynamical
model of quantum mechanics (Pardy, 2001d) the classical terms from the
hydrodynamics involving the viscosity properties of a fluid, then we
get the following evidently the beautiful quantum Navier-Stokes equations:

$$m\left(\frac{\partial {\bf v}}{\partial t} + ({\bf v}\cdot \nabla){\bf v}
\right) = - m \nabla V_{q} - \frac{1}{n}\Delta p  +
\frac{\eta}{n}\Delta{\bf v} + \frac{1}{n}\left(\xi + \frac{\eta}{3}\right)
{\rm grad}\;{\rm div}\;{\bf v},\eqno(15)$$
where $V_{q}$ is the known quantum mechanical potential:

$$V_{q} = - \frac{\hbar^{2}}{2m}\frac{\Delta\sqrt{n}}{\sqrt{n}}.\eqno(16)$$
and potential $V$ is replaced by the macroscopic pressure $p$.

This equation (15) can be considered as the quantum mechanical
model of the quantum process involving
dissipation. We hope that this equation will be confirmed by
experiment and then it will be of the same value as the original
Schr\"{o}dinger equation of the quantum mechanics.

The second possibility of the method of playing with equation is the
equation of the so called quantum magneto-hydrodynamics which can be
eassily obtained from the modification of the quantum hydrodynamical 
equation(Madelung, 1926)
inserting in it the magneto-hydrodynamical term. The result is the
following equation:

$$m\left(\frac{\partial {\bf v}}{\partial t} + ({\bf v}\cdot \nabla){\bf v}
\right) = - m \nabla V_{q} - \frac{1}{n} \nabla p
+ \frac{1}{4\pi n}({\rm rot}{\bf H}\times{\bf H}),
\eqno(17)$$
where
and ${\bf H}$ is the vector of the magnetic field. The last equation
involves quantum corrections to the classical magneto-hydrodynamical
equation. The Alfv\'en waves follows for instance form the
magneto-hydrodynamical equation. It means that equation (17) involves
the quantum corrections to the Alfv\'en waves.

The dissipative terms can
be evidently inserted in the equation (17) in order to get quantum
Navier-Stokes magneto-hydrodynamical equation.

$$m\left(\frac{\partial {\bf v}}{\partial t} + ({\bf v}\cdot \nabla){\bf v}
\right) = - m \nabla V_{q} - \frac{1}{n} \nabla p \quad + $$

$$\frac{\eta}{n}\Delta{\bf v} + \frac{1}{n}\left(\xi + \frac{\eta}{3}\right)
{\rm grad}\;{\rm div}\;{\bf v}
+ \frac{1}{4\pi n}({\rm rot}{\bf H}\times{\bf H}).
\eqno(18)$$

Now, let us consider the so called nonlinear Schr\"odinger equation and
its solution. It was shown that it was also derived by the method of
playing with equation (Pardy, 2001d).

The motivation for
introduction of the nonlinear quantum theory was in expecting
of the better understanding of the synergism of waves
and particles. Bia{\l}ynicky-Birula and
Mycielski (1976) considered the generalized Schr\"{o}dinger
equation with the additional term  $F(|\Psi|^2)\*\Psi$, where
$F$ was some arbitrary function which they later specified
to be $-b(\ln|\Psi|^2)$ as a consequence of the factorization
of the wave function for the composed systems. The
same equation was also obtained by Lemos (1983) within
the context of the stochastic quantum mechanics. Using the method of playing
with equation, we derived the generalized  Schr\"{o}dinger equation
as follows (Pardy, 2001d).

$$ i\hbar\*\frac {\partial\*\Psi}{\partial\*t} = - \frac {\hbar^2}{2m}\*
\Delta\*\Psi + V\*\Psi - b(\ln|\Psi|^2)\*\Psi,\eqno(19)$$

\noindent
where $b$ is positive and the nonlinear term contents
the unit constant with the dimensionality of (length)$^3$.

It is easy to show that eq. (19) follows
from the hydrodynamical model of quantum mechanics (Bohm et al., 1954;
Madelung, 1926; Rosen, 1974) after
adding into the Euler hydrodynamical equation
the pressure $-b\*|\Psi|^2$ and by the inverse procedure to
the derivation of the hydrodynamical model (Pardy, 2001d).

The solution of eq. (19) can be assumed in the soliton-wave form
(in the one-dimensional case with $V=0$) (Pardy, 1994a; Pardy, 2001d):

$$ \Psi(x,t)= c\*G(x-vt)\*e^{i(kx-\omega\*t)},\eqno(20)$$

\noindent
where $c, v, k $ and $\omega$ are real numbers. After insertion
of function (20) into eq. (19), we get

$$
\Psi(x,t) = c\*e^{a/B}\*\exp\left\{-\left[\frac
    {1}{4}\*B\*(x-vt+d)^2\right]\right\}\*e^{i(kx-\omega\*t)}, \eqno(21)$$

\noindent
where

$$v= \frac{\hbar\*k}{m}, B  = \frac {4mb}{\hbar^2},\eqno(22)$$

\noindent
and

$$a = \frac {B}{2} - \frac {2m}{\hbar}\*\omega + k^2 -
\frac {2m}{\hbar^2}\*b\*\ln\*c^2,
\eqno(23)$$

\noindent
with $d$ being some constant. The constant $c$ can be
determined from the normalization condition

$$\int_{-\infty}^{\infty}\;\Psi^{*}\Psi\;dx = 1.
\eqno(24)$$

The probability density is $\delta_{m}(\xi) = \Psi^{*} \Psi, $ and from
eq. (21) we get (for $d=0$)

$$\delta_{m}(\xi) = \sqrt{\frac {m\alpha}{\pi}}\*e^{-\alpha\*m\xi^2},\quad
\xi = x-vt, \quad
\alpha = \frac {2b}{\hbar^2}.
\eqno(25)$$

It is obvious that $\delta_{m}(\xi) $ is the delta-generation
function, which tends for $ m \rightarrow \infty $ just to the Dirac
delta function. The advantage of the $\delta_{m}$ function is in its behavior
for the sufficiently the large mass $m$, because in this case it describes
the strongly localized motion of a particle. In other words, function
$\delta_{m}$ describes the classical motion of a particle with large
mass $m$ in contrast to the standard quantum mechanics. Such behavior
obviously gives the logical support for the competence of the
logarithmic quantum mechanics.

Other interesting feature of the nonlinear QM consists in fact that
it solves the Schr{\"o}dinger cat as a consequence of the broken principle
of superposition (Pardy, 2001d).

If the nonlinearity in the Schr\"o dinger equation is of the
quadratic form (Pardy, 1989b), we get so called Gross-Pitaevskii equation

$$i\hbar\*\frac {\partial\*\Psi}{\partial\*t} = - \frac {\hbar^2}{2m}\*
\Delta\*\Psi + V\*\Psi - \alpha|\Psi|^2 \*\Psi,\eqno(26)$$
which describes the quantum motion of superfluid described in many
articles including author's one (Pardy, 1989b).

\section{The problem of understanding of mathematics and physics}
\hspace{3ex}
The high creativity in mathematics, physics, or in other sciences
is based on the deep understanding of the content and techniques
used. The creativity without understanding is a free play with the
mathematical or theoretical objects, but giving nothing interesting.
Only understanding of the scientific disciplines gives the
interesting and valuable results. However, what is the understanding
as such? Especially what is the understanding of mathematics or
theoretical physics? Let us try to answer this question.

Understanding in science, especially in physics, involves the
{\bf knowledge of explanation
of a fact, prediction of new consequences of it  and knowledge of
possible applications.}

Paul Ehrenfest (Weisskopf, 1971):

\begin{quote}
{\bf ``refused to admit that something is understud if one understands
 only the mathematical derivation''}
\end{quote}
Or, another statement of Ehrenfest (Weisskopf, 1971):

\begin{quote}
{\bf "Physics is simple but subtle"}
\end{quote}

In case we consider some physical
effect, the understanding of this effect means how does the effect
arise, what are the general causes of it and how it is related to the
other part of nature.

Let us imagine a long series of syllogisms with components A, B, C,
..., in which the conclusions of
those that precede form the premises of those that follow. Or,
It is obvious that everybody intelligent person is able to grasp the
syllogism in the isolated form, or, in the form
$A  \Longrightarrow  B;\quad B  \Longrightarrow  C;\quad
C  \Longrightarrow  D ;\quad ...   \Longrightarrow  END $.

However, between the moment when a man meets the first proposition
and the END much time will sometime  have elapsed and it may well
happen that a man has been influenced by some unexpectable events
which change the meaning of the individual steps
and also the meaning of the whole chain. In such a way the chain of
syllogisms forms no harmonic system for a man and the subjective
reflection of such situation is the feeling of some misunderstanding.
In other words understanding supposes not only knowledge and
understanding of the individual steps but the simultaneous retaining
them in the memory in the original chain. From these words it follows
that people who have weak mathematical memory are not able to grasp
the total chain of the mathematical steps and therefore are not able
to understand mathematics.
On the other hand the computer calculations
inform us that the true results can be obtained without understanding
in the human sense because computer works algorithmically
without understanding.

After analysis of the notion {\bf understanding} we have found
miscellaneous view on the notion of understanding. For instance:

\begin{quote}
{\bf a) understanding means to see the premises from which the theory
arises, what are the fundamental results of the theory and the method used\\
b) understanding means the knowledge of logical systematization\\
c) understanding is the knowledge of the model of the system or situation \\
d) understanding means the knowledge of the operational system of the
 theory and its physical meaning\\
e) understanding is process which is tested by
 the ability of solving physical problems of given theory\\
f) understanding means ability of formulation of the new problems\\
g) understanding means the knowledge of all questions formulated in the
boundaries of the given theory}
\end{quote}
and so on. It is obvious that it is
possible to create other definitions of understanding which stress
specific aspects of process of understanding.

Understanding can be tested by {\bf explanation}. What is explanation?
It is the verbal process which can be no more then
a {\bf description which simplify.} It describes situation
in terms of fundamental
factors, being so fundamental that no description of them is possible.
The fundamental factors called "absolute" may be given names and the
number of them should be a minimum. The explanation involves
understanding and should employ only such conceptions which are in common
use. It is essential that the explanation must be logically correct
and actually true. It is well known from the pedagogical practice that the
explanation of the unknown object can be realized easily by {\bf
exemplification.}

According to Poincar\'{e} (1913),
understanding in mathematics appears as
sudden illumination. However, it is not absolutely spontaneous, but
appears as a result of the {\bf long course of the previous work.}

The integral part of understanding is the feeling of harmony which
form a internal signal informing us on understanding of the mathematical or
the physical system of ideas. If a man feels during the cognitive
process certain disharmony, then this is the signal informing him on
the necessity to analyze the problem once again.

At present time, {\bf the time of the artificial intelligence},
the question arises how to realize explanation with so called computer
understanding of the problem situation,
and how to make predictions by computer by pure
deductive algorithm. For prediction, it is the {\bf logical process which
is equivalent to mathematical deduction}. Inductive predictions are
not valid because the results of induction are hypotheses which must
be proved.
We belief that it is possible to solve
the general problems by computer and also to produce problems by computer.

\subsection{Understanding of relativity}

The special theory of relativity has familiarized
us with the notion of thinking of space-time in terms of geometry.
In other words we describe space-time as a differential manifold
endowed with the Minkowski metric

$$ds^2 = c^2dt^2 - (dx^2 +dy^2 +dz^2).\eqno(27)$$

Putting

$$t = t(x_1, x_2, x_3, x_4); x = x(x_1, x_2, x_3, x_4);
y = y(x_1, x_2, x_3, x_4); z = z(x_1, x_2, x_3, x_4),\eqno (28)$$
we get after insertion of the transformations (28)
into the space-time element (27):

$$ ds^2 = g^{\mu\nu}dx_{\mu}dx_{\nu},\eqno(29)$$
which is equivalent to the original element (27). The  coefficients
${g^{\mu\nu}}$ form the so called metric tensor. It is determined
unambiguously by the transformations (28). It is
possible to show that if we choose the metric tensor arbitrarily,
then there is no transformations of the form (28), from
which the coefficients ${g^{\mu\nu}}$ follow. The form (29) in
this case defines so called Riemann geometry. The Einstein idea
consists in the assumption that the space-time in the presence of the
gravitational field is not Euclidean but Riemannian and that the
motion of a body in the gravitational field is equivalent to the
motion of the same body in the Riemann space-time along the geodesic
path which is described by the equation

$$\frac {d^2x^{\mu}}{ds^2} = \Gamma_{\alpha\beta}^{\mu}\frac {dx^{\alpha}}{ds}
\frac {dx^{\beta}}{ds}, \eqno(30)$$
where $\Gamma_{\alpha\beta}^{\mu}$ are so called the Christoffel
symbols and they are dependent on $g^{\mu\nu}$. The geodesic path can
be defined also mechanically in the 2D-geometry as a trajectory of the
infinitesimal roller skate moving on the 2D-surface. The
definition of the mechanical infinitesimal roller skate
in the 4D-surface was not given till present time.

Why did Einstein used the Riemann geometry and not the different
approach? From the history of mathematics it is known that
Gauss (1777-1855), and also Loba\v{c}evskii (1792-1850)considered
space and time as the objective reality and created the
philosophical idea that the geometry of space is determined
by the physical properties of it. Or, it may not be necessary
Euclidean.  Later Riemann (1826-1866) created geometry of his name.
So the idea that space, or, space-time can be
curved was not new and Einstein was free to accept the physical
idea that the curvature of space-time is caused by gravity.

In case that the transformations (28) do not involve time,
we get eq. (30) in the three-dimensional form with
$ds = dt$ and then they are equivalent to the Lagrange equations. They
are involved in the every textbooks on the tensor calculus.

The metric tensor $g^{\mu\nu}$ in the general theory of relativity and
gravitation can be determined by solution of
the Einstein equation

$$R^{\mu\nu} -\frac{1}{2}g^{\mu\nu}R = (8{\pi}G/c^4)T^{\mu\nu}.
\eqno(31)$$
where $G$ is the Newton gravitational constant, $c$ is the
velocity of light and $T^{\mu\nu}$ is the tensor of the energy and
momentum of the physical system and $R^{\mu\nu}$ is the Ricci
tensor. Let us remark only, that while Einstein derived equations by
the very sophistic operations involving also the so called
fictitious experiments with elevator, David Hilbert derived these
equations easily and elegantly from the variational principle
postulating only the appropriate Lagrangian and proved that
general theory of relativity is only one of the
problems of the Lagrangian mechanics. {\bf So, knowledge of
mathematics is one of the most practical thing in constructing physical
equations.}

The fundamental objections to the Einstein approach is that the 
three-dimensional
space cannot be curved in the three dimensional space and we have no evidence
on the more than three-dimensional space. The objection is serious because
the internal geometry on the sphere is possible only as the existence
of the four-dimensional space in which the sphere is immersed.
The resolution of this paradox is possible using the optical analogy.
In optics there exist so called the optical length which is in no case
equivalent to the geometrical length. In the nonhomogenous optical
medium the distance of two bodies A and B is given
by the optical length, or by the geometrical length which is always
Euclidean. On the other hand in case of the space time with the gravitational
field there exist, according to Einstein, only the physical length
which can be measured only by light rays and clocks. Because the
experimentalist is immersed in the Riemann space time, he has no possibility
in this space-time to define the Euclidean length in contradiction
to the optical situation.

Let us remark also that there many missing problems and missing
solutions in general
relativity. For instance, the solution of the mathematical pendulum,
the Foucalt pendulum, physical pendulum, free fall, the special
relativistic limit of general relativity and so on. Some of these
problems are discussed by Santilli (2005).

Now, the question arises, if the Einstein approach is the sole method
how to describe gravity. Poincar\'{e} claims that the real objects
are not described only by geometry $G$ , but by the
geometry together with the physical law $P$, or by $G + P$. As
the geometry is created by the a' priori way, the physical law must be
added to such a' priory geometry in order to get the appropriate
description of reality.

This alternative way was realized for instance by  Gupta
(1952), Feynman (1963) and  Schwinger (1976) who has
replaced the Einstein theory by the quantum field theory of
gravitation where gravity is described by the tensor of massless field of the
second rank which corresponds to the bosons with zero mass, having the
spin $2$ and helicity $\pm 2$.  Attractive forces between bodies and
all other effects are in this theory caused just by these bosons.

The basic formula in the source
theory is the vacuum-to-vacuum amplitude (Schwinger, 1976):

$$\langle 0_{+}|0_{-} \rangle = e^{\frac
  {i}{\hbar}\*W(S)},\eqno(32)$$
where the minus and plus tags on the vacuum symbol are causal
labels, referring to any time before and after space-time region
where sources are manipulated. The exponential form is introduced
with regard to the existence of the physically independent
experimental arrangements which has a simple consequence that the
associated probability amplitudes multiply and corresponding
$W$ expressions add. (Schwinger, 1976).

In the flat space-time the field of gravitons is described by
the amplitude (32) with the action ($c = 1$):

$$W(T) = 4\pi\*G\*\int (dx)(dx')\Bigl[T^{\mu\nu}(x)
\*D_{+}(x-x')T_{\mu\nu}(x')
 -  \frac{1}{2}\*T(x)D_{+}(x-x')T(x')\Bigr],\eqno(33)$$
where the dimensionality of $W(T)$ is the same as the dimensionality
of the Planck constant $\hbar$. $T_{\mu\nu}$ is the tensor of momentum and
energy, and for particle moving along the trajectory
${\mbox {\bf x}} = {\mbox {\bf x}(t)}$
it is defined by the equation

$$T^{\mu\nu}(x) = \frac{p^{\mu}p^{\nu}}{E}\*
\delta({\mbox {\bf x}} - {\mbox {\bf x}}(t)), \eqno(34)$$
where $p^{\mu}$ is the relativistic four-momentum of a particle
with a rest mass $m$.

The last formulas was used to determine the production of gravitons by the
binary system  (Pardy, 1983b, 1994d). The method of calculation of the
emission of gravitons is an analogue of the method of calculation of the
photon emission by the binary system 
(Pardy, 2000a, 2000b, 2001c, 2002b, 2004d).

The special synthesis of the curvature approach and the spin 2
approach enabled to determine the so called
gravitational \v Cerenkov radiation which is analogous to the \v Cerenkov
radiation in a dielectric medium (Pardy, 1983a, 1983c, 1989a).
Calculation was applied with radiative corrections (Pardy, 1984; 1994c, 1994e)
to zero temperature and finite temperature situation
(Pardy, 1994b, 1995).
This effect was still not observed
similarly as the Hawking effect. While the
electrodynamical \v Cerenkov effect with massive photons was calculated
(Pardy, 2002a) together with radiative corrections,
it is not excluded that the gravitational \v Cerenkov effect with
massive gravitons is of the physical meaning. The solution of this
problem was still not published. The gravitons can be evidently produced by
plasma fluctuations. This problem was not solved by author. The first step
was performed in the case of photons generated by the plasma fluctuations
(Pardy, 1994f).

Gravity as every physical theory is dependent on crucial experiments.
For instance, on the deflection of light. It was confirmed by experiment.
However if photon is massive, or if it is considered as a quantum particle,
then it is necessary to make new analysis (Pardy, 2001a, 2005a).

We can say that in case of the general relativity
the understanding of it is elementary if we accept the Hilbert
variational approach and  the curved space-time as physically meaningful.
The existence of the massive gravity is possible  from the postulation of
the massive gravitons. The existence of the massive classical gravity
is open question and it is open problem in the definition of the
massive classical gravitational field.

While the Pythagoras theorem $c^2 = a^2 + b^2$ is valid for the right angle
triangle in the Euclidean geometry, it si nor valid on the
spherical surface and on the general
Riemann surface. Similarly the plane geometry Heron formula

$$ P = \frac{1}{2}\sqrt{s(s-a)(s-b)(s-c)};\quad
s = \frac{1}{2}(a+b+c)\eqno(35)$$
is not valid on the spherical surface. It means that spherical
geometry, or Riemann geometry is substantially different with regard
to flat geometry. It also means that Einstein theory of Riemann
space-time substantially differs from the flat space-time
geometry. The experiment must decide if the Einstein theory is more
fundamental than the Schwinger source theory of gravity.

It is surprising that in the Newton period Pierre Gassendi developed the
so called string model of gravity which states that between body A and
B there is a string which transmits the attractive force. It seems
that Faraday used the Gassendi ideas in his theory of
electromagnetism. The string model of gravity is now considered as the
problem of mathematical physics and it is not excluded that theory of
this model will give fresh ideas in the string theory of matter
(Pardy, 1996a; 2005b).

Definition of the space-time element and the Einstein equation with
the cosmological constant
makes it possible to define also the cosmological dislocations (Pardy, 2005a)
and antidislocations. The definition of these dislocation in
the flat space-time quantum gravity is open problem.
There is no experimental confirmation of such dislocations till
present time. Dislocations and antidislocations can annihilate to form
light bursts in universe (Pardy, 2005a).

While Einstein used the fictitious experiment with accelerated elevator,
there is a different approach to description of
accelerated systems. The nonrelativistic approach
by Pardy (2004a),  and the relativistic discussion by Friedman
and Gofman (2005).

In the nonrelativistic approach, the
two systems $S(0, x, y, z)$ and  $S'(0, x', y', z')$ are considered,
where system $S'$ moves in
such a way that $x$-axes converge, while $y$ and $z$-axes run parallel and at
time $t = t' = 0$ for the beginning of the systems $O$ and $O'$ it is $O
\equiv O'$ (Pardy, 2004a).

Let us suppose that system $S'$ moves relative to some basic
system $B$ with acceleration $a/2$ and system $S'$ moves relative to system $B$
with acceleration $-a/2$. It means that both systems moves in the
opposite direction one another
with acceleration $a$ and they are equivalent because
in every system it is possibly to observe the unit force caused by the
acceleration $a/2$. In other words no system is inertial.

Now, let us consider the formal transformation equations between two systems.
At this moment we give to this transform only formal meaning because at this
time, the physical meaning of such transformation is not known.
On the other hand,
the consequences of the transformation will be shown very interesting.

We write the transformation equations in the form (Pardy, 2004a):

$$x' = a_{1}x + a_{2}t^{2}, \quad y' = y,\quad z' = z,
\quad t' = \sqrt{b_{1}x + b_{2}t^{2}},\eqno(36)$$
where constants involved in the equations will be determined from the
viewpoint of kinematics. Since from the viewpoint of kinematics, both systems
are equivalent, for the inverse transformation to the transformation (36)
it must hold:

$$x = a_{1}x' - a_{2}t'^{2}, \quad y' = y,\quad z' = z,
\quad t = \sqrt{-b_{1}x' + b_{2}t'^{2}}.\eqno(37)$$

The minus sign with coefficients $a_{2}$ and  $b_{1}$ appearing for the reason
that constant $a_{2}$ has the rate of acceleration while constant $b_{1}$
the rate of inverse value of acceleration.

Similarly, as in inertial systems, the hypothetical requirement can be now
expressed that the transformation equations for system moving relative to
themselves with acceleration include a suitable invariant function. Let us now
define such transformations as follows:

$$x = \frac {1}{2}\alpha t^{2}; \quad x' = \frac {1}{2}\alpha t'^{2},
\eqno(38)$$
where $\alpha$ is the constant having the rate of acceleration.

If we now substitute (37) into (36), we obtain
from $[x' = (1/2)\alpha t'^{2}] \Longleftrightarrow [x = (1/2)\alpha t^{2}]$

$$x' = \Gamma(a)(x - \frac {1}{2}at^{2}), \quad t'^{2} =
\Gamma(a)\left(t^{2} - \frac{2a}{\alpha^{2}}x\right) ;\quad
\Gamma(a) = \frac {1}{\sqrt{1 - \frac {a^{2}}{\alpha^{2}}}}\eqno(39)$$
and $ \quad y' = y,\quad z' = z,$.

Let us remark that the more simple derivation of the last transformation can be
obtained if we perform in the Lorentz transformation the elementary
change of variables as follows:
$t \rightarrow t^{2},\quad t' \rightarrow t'^{2}, \quad  v \rightarrow
\frac {1}{2}a, \quad c \rightarrow \frac {1}{2}\alpha$.

The physical interpretation of this nonlinear transformations is the
same as in the case of the Lorentz transformation only the physical
interpretation of the invariant function $x = (1/2)\alpha t^{2}$ is
open. The time transformation requires the correct interpretation of
time. We use the time delay here as in the Fok book, where Fok
writes that time delay is the form of the measurement procedure
(Fok, 1961). It was proved that Fok interpretation of time is
correct (Pardy, 1969). Some relativists suggest different
interpretation of time dilatation.

Similarly, we can consider the
length contraction in special theory of
relativity (STR). The magnitude of the moving length depends on the
measurement  procedure. Let us prove it.
The rod velocity $v$ does not depend on the light
velocity. Then, if the end points of rod  are $A$ and $B$, and, if we define
the point of observation $O$ in $S$ , and, if the points A and B
are identical with point $O$ at time $t_{A}, t_{B}$, then the length of
the moving rod is evidently $L = v (|t_{A} -  t_{B}|)$. This
definition of length as dual to the contracted length in STR and it is missing
definition in STR. The measurement of the length by the \v Cerenkov effect
was published recently (Pardy, 1997a).

If we want to measure the laser pulse, then it is necessary to use the
more sophistic approach. The duration of the laser pulse
$\Delta t$ of the  rest laser
in the system $S$ follows from the relation $c\Delta t =  n\lambda$
where $\lambda$ is the wave length of the laser light and $n$ is the
number of waves. Observer in the system $S'$ which moves at the
velocity $v$ with regard to the laser system $S$ writes
$c(\Delta t)' = n\lambda'$ where $\lambda'$ is transformed wave length
according to special relativity theory. Or,

$$\lambda' = \lambda\sqrt{\frac{1 - v/c}{1 + v/c}}\eqno(40)$$
for the observer moving toward the laser pulse. This relation, known as
the relativistic Doppler formula can be
used as a test of special theory of relativity to show that the
one-way velocity of light is $c$. The nonrelativistic relation is
$\lambda' = \lambda(1 - v/c)$ and the nonrelativistic $(\Delta t)'$
differs from the relativistic $(\Delta t)$.

We know from the history of physics, that Lorentz
transformation was taken first as physically meaningless by Lorentz
himself and later only Einstein decided to put the physical meaning
to this transformation and to the invariant function $x = ct$.

The transformation (39) forms
one-parametric group with parameter $a$. To prove it we must prove
by the direct calculations the four requirements
involving in the definition of a group. However, we know,
that using relations $t \rightarrow t^{2},\quad t' \rightarrow t'^{2},
\quad  v \rightarrow
\frac {1}{2}a, \quad c \rightarrow \frac {1}{2}\alpha$, the nonlinear
transformation is expressed as the Lorentz transformation
forming the one-parametric group. And this is a proof.
Such proof is equivalent to the proof by direct calculation.
The integral part of the group properties is the so called
addition theorem for acceleration.

$$w_{3} = \frac {w_{1} + w_{2}}{1 + \frac {w_{1}w_{2}}{\alpha^{2}}}
\eqno(41)$$
where $w_{1}$ is the acceleration of the $S'$ with regard to the system $S$,
$w_{2}$ is the acceleration of the system $S''$ with regard to the system
$S'$ and $w_{3}$ is the acceleration of the system $S''$  with regard to the
system $S$.

The relation (41),
expresses the law of acceleration addition theorem
on the understanding that the
events are marked according to the relation (39). In this formula as well as
in the transformation equation (39) appears constant $\alpha$ which cannot be
calculated from the theoretical considerations, or, from the theory.
What is its magnitude and whether there exists such a physical field that is
consistent with the designation of the events given by the relations (39) can
be established only by  experiments. On the other hand the constant
$\alpha$ has physical meaning of the maximal acceleration and its meaning
is similar to the maximal velocity $c$ in special relativity.

It is not excluded that the maximal acceleration constant
determines the upper limit of the Higgs boson and links
the mass of $W$ boson with the mass of the Higgs boson.
The LHC and HERA experiments will give the answer to this problem.
At the same time it is not excluded that the maximal acceleration constant is
related to the so called Hagedorn temperature which is the phase boundary
temperature between the hadron gas phase and the deconfinement state of mobile
quarks and gluons. In other words, acceleration of quarks and gluons in the
hadron gas is restricted by the maximal acceleration.

We can also suppose in analogy with the special relativity
that mass depends on the acceleration for small
velocities,
in the similar way as it depends on velocity in case of uniform motion.
Of course such assumption must be experimentally verified and in no case
it follows from special theory of relativity, or, general theory of
relativity. So, we postulate
ad hoc, in analogy with special theory of relativity:

$$m(a) = \frac {m}{\sqrt{1 - \frac {a^{2}}{\alpha^{2}}}};
\quad v \ll c, \quad a = \frac{dv}{dt}.\eqno(42)$$

Let us derive as an example the law of
motion when the constant force $F$  acts on the body with the rest
mass $m$. Then, the Newton law reads [15]:

$$F = \frac{dp}{dt} =
m\frac{d}{dt}\frac {v}{\sqrt{1 - \frac {a^{2}}{\alpha^{2}}}}.
\eqno(43)$$

The solution of this equation was found (Pardy, 2003b, 2004a).
For F $\rightarrow \infty$, we get $v = \alpha^{2}t$. This relation can
play substantial role at the beginning of the Big Bang, where the accelerating
forces can be considered as infinite, however the law of acceleration
has finite nonsingular form.

We know that the dependence of the mass on acceleration follows from
quantum electrodynamics (Ritus, 1979). Our case is based
on the different approach.
In case that the constant ``maximal acceleration'' will be confirmed for
instance by LHC, then, the high energy physical theories, gravitation
and cosmology must be modified, including the Hawking effect. Evidently,
this constant will generate revolution in cosmology.

\subsection{Understanding of quantum mechanics}

Although the mathematical structure of quantum mechanics is
elementary, or, in other words, this theory is only
Schr\"{o}dinger equation applied to many
physical situation, the most difficult problem is the derivation of
the Schr\"odinger equation and the origin of
probability behavior of an electron. Till  present time, the problem was
not solved.

To solve this problem,
Feynman invented so called path integral approach to quantum mechanics
with the goal to remove the mystery of quantum mechanics.
The substance of the Feynman path integral
is, that the wave function of an electron ,
is the sum amplitudes taken over all classical trajectories of a particle
(Blokhintzev, 1971;  Pardy, 1973c).
The  interpretation of the Feynman sum is a such,
that electron moves from point A to B along all classical trajectories
connecting A and B (Pardy, 1973c).

According
to Feynman, the total probability amplitude $U(A, B)$ from a particle
transition from point A to point B is as follows

$$ U(A, B) = \frac{1}{C}\sum_{\stackrel{over\; all \; trajectories}
{from \;A\; to\; B}}
 e^{\frac {i}{\hbar}S[x(t)]}, \eqno(44)$$
where $S$ is the classical Hamilton-Jacobi action function

$$S = \int_{t_{A}}^{t_{B}}L[{\dot x}(t); x(t)]dt, \eqno(45)$$
where $L$ is Lagrange function.

We can see that all amplitudes $\exp\{(i/\hbar)S\}$ are
multiplied only by a constant
$1/C$. In
case that the electron is in some stochastic medium, for instance
inside the black-body where the thermal stochastic photons are present
the sum must be modified by some weight function. We take the Wiener
measure of the Brownian motion for the weight function
of a particle in a stochastic medium (Pardy, 1973c).
In other words, the original trajectories are modified.
If the
result gives the physical meaningful consequences, then also the
modification of trajectories is physically meaningful.
It was shown that the result of such
generalization is the change of mass of electron according to the
formula (Pardy, 1973c):

$$m \quad \rightarrow \quad m + i\delta, \eqno(46)$$
where $\delta$ is related to the so called diffusion constant of 
the Brownian motion. The complex
mass of electron means that no energetic level is stationary, which is
very natural when electron is moving in the stochastic medium. However,
if the space as such is not euclidean but stochastic, or,
infinitesimally deformed, for instance in the vicinity of the thermal
Big Bang, then it
seems that the modification of the trajectories is mathematically
necessary. Hawking (1989) accepted the Feynman theory over the path with the
interpretation that electron moves from A to B by every possible
path. The possible modification of the Feynman integral by some
additional properties of space time is not involved in the Hawking
monograph. Similarly, such approach to the Feynman integrals
was not elaborated in the literature.

On the other hand the complex mass of electron naturally appears
in the quantum electrodynamical description of the synchrotron
radiation and it is possible  from it to calculate the spectrum of the
synchrotron radiation. The nonstability by the complex mass,
of the orbit of electron in
synchrotron was discussed (Pardy, 1985).

There is one specific case where Feynman sum is restricted to one
trajectorie. This is so called Volkov solution of the Dirac
equation. The solution is of the form (Berestezkii, 1989;
Pardy, 2003a, 2004b, 2004c):

$$\psi_{p} = R \frac {u}{\sqrt{2p_{0}}}e^{iS}  =
\left[1 + \frac {e}{2kp}(\gamma k)(\gamma A)\right]
\frac {u}{\sqrt{2p_{0}}}e^{iS},
\eqno(47)$$
where $u$ is an electron bispinor of the corresponding Dirac equation

$$(\gamma p - m)u = 0,\eqno(48)$$
with the normalization condition ${\bar u}u = 2m$

The mathematical object $S$ is the classical Hamilton-Jacobi function,
which  was determined in the form:

$$S = -px - \int_{0}^{kx}\frac {e}{(kp)}\left[(pA) - \frac {e}{2}
(A)^{2}\right]d\varphi \eqno(49)$$
and $A(\varphi)$ is a periodic  potential with $\varphi = kx$. For
the electron motion in the two potentials  $A(\varphi)$ and  $B(\chi)$,
and for the specific configuration of fields $A, B$ (Pardy, 2004c), we
get

$$\psi = R(\varphi)R(\chi)\frac {u}{\sqrt{2p_{0}}}
e^{iS(\varphi)}e^{iS(\chi)}. \eqno(50)$$

The Volkov wave function is quantum mechanical one and it
involves classical action, however no
integral over trajectories is present and at the same time
this is not the WKB
approximation. It means that all trajectories are mutually canceled
excepting the classical one. Why? This is an open and the
missing problem in the monographs on Feynman integral.

\section{The selection of facts}
\hspace{3ex}

It is well known that {\bf the explosive development of industry was
iniciated in the beginning by the absolutely nonpractical ideas (The
maxwell equations, the Hertz experiments with the electromagnetic
waves, the special theory of relativity, the Boole algebra and the
mathematical logic and so on)}. Some of the initiators of such
unpractical ideas were unpractical people who died poor and who never
thought of the practical employment of their ideas. Nevertheless,
they had a guide that was not their own caprice. It was the
happiness for society that such ideas were not refused. Of course,
some ideas are always rejected and {\bf the question arises what are the
mathematical and physical ideas which influence the life of the
industrial society by the maximal way}.

According to Poincar\'{e} (1913), {\bf the most
important facts are those which can be used several times, or which
have a chance to recurring}. However, what are the facts of such
properties? According to Poincar\'{e} (1913), the simple
facts. Because
it is evident that a complex fact is composed from many simple facts
by chance and that only a still more probable chance could ever so
unite them again, on the contrary to the case of a simple fact.

The question is what is the simple fact. The astronomer has found the
simple fact a point which replaces the universe bodies, because the
distance between the stars are immense, that each of the universe
body appears only as a point and qualitative differences disappear and
because a point is a simpler object than a body with shape and qualities.

Particle physicists has found that the simple facts in the microworld
are quarks which are point-like, massive and charged together with
electrons, positrons, $\mu$-mesons , $\tau$-mesons and neutrinos which are also
(approximately) point-like with regard to the present
experimental knowledge. All
matter is composed from these particles and from intermediate bosons
which cause the electromagnetic, weak, strong and gravitational
interactions between particles and bodies.

The similar situation is in biology, chemistry and so on, where it is
possible to establish the simple facts.

The fundamental rule to perform a good selection of facts is not
ascertain resemblances and differences of the mathematical objects,
however, {\bf to discover similarities hidden under apparent
discrepancies}. This is what gives value to certain facts that come to
complete a whole and shows that it is the faithful image of other
known wholes. Mathematicians and physicists do not select facts
randomly. They try to condensate a great deal of experience and great
deal of thought into a small volume and that is why a little book on
mathematics and theoretical physics contains so many pieces of knowledge.

According to Poincar\'{e} (1913), the selection of basic facts and
searching for these facts is accompanied by the specific mathematical
emotions which can be called as {\bf the emotion of the mathematical
beauty}. These emotions for the motivation for study of mathematics
and theoretical physics. Poincar\'{e} (Poincar\'e, 1913; Wolpert, 1993) writes:

\begin{quote}
{\bf "The scientist does
not study nature because it is useful; he studies it because
he delights in it, and because it is beautiful.
Of course, I do not speak here of that beauty which strike the senses, the
beauty of qualities and appearance; not that I undervalue such beauty, far
from it but it has nothing to do with science. I mean the profound beauty
which comes from the harmonious order of the parts and which a pure
intelligence can grasp."}
\end{quote}
Or,

\begin{quote}
{\bf "If nature were not beautiful, it would not be worth knowing
and life would not be worth living".}
\end{quote}

The mathematical beauty is not the beauty which strikes senses, the
beauty of qualities and appearance. This is (Poincar\'e, 1913)

\begin{quote}
{\bf ``intimate beauty which
comes from the harmonious order of its parts, and which a pure
intelligence can grasp"} (Poincar\'e, 1913).
\end{quote}

Or, (Poincar\'e, 1913)
\begin{quote}
{\bf  `` ... the care for the beautiful leads us to the some selection as
care for the useful. Similarly, economy of thought, that economy of
effort, which according to Mach, is the constant tendency of science,
is a source of beauty as well as a practical advantage"}.
\end{quote}

The question arises, whether the mathematical beauty is the
sufficient condition for the theory to be physically true. It
is necessary to say that it is possible to invent many mathematical
theories, equations, relations and so on which have mathematical
beauty but which do not correspond to nature.

We known that the basic fact of Newton physics is the notion of
force. {\bf Every force is
invisible,} including Lorentz force and forces in particle physics.
They transform according to
the transformation laws of the special theory of relativity.
The basic theoretical object in the Lagrange mechanics is
the Lagrange function $L = T - V$, where $T$ is the kinetic energy of
a particle and $V$ is the potential energy. Potential energy was introduced by
Rankine -- the missing information in textbooks. Force in
Lagrange mechanics is defined from potential. Schr\"{o}dinger
equation involve potential and not force. Similarly the general
relativity by  Einstein does not work with force. Force in quantum
field theory is derived from the interactions. Potential energy
is derived from the Green function of a particle. While force and
potential are  basic ingredients in classical mechanics,
in quantum field theory not. We use here the propagator, or the Green function.

The propagator, or the Green function,
for the massive photon in dielectric medium with index of refraction
$n$ was derived as (Pardy, 2002a)

$$D_{+}(x-x',m^{2}) = \frac{i}{c}\frac{1}{4\pi^2}
\int_{0}^{\infty}\,d\omega\,\frac{\sin[\frac{n^2\omega^2}{c^2}-
\frac {m^2\*c^2}{\hbar^{2}}]^{1/2}
|{\bf x}-{\bf x}'|}{|{\bf x}-{\bf x}'|}\*e^{-i\omega\*|t-t'|}.
\eqno(51)$$

The potential is
according to Schwinger defined by the formula:

$$V({\bf x} - {\bf x'}) = \int_{-\infty}^{\infty}d\tau
D_{+}({\bf x - \bf x}',\tau). \eqno(52)$$

The $\tau$-integral can be evaluated, in case of the massless photon,
using the mathematical formula

$$\int_{-\infty}^{\infty}\,d\tau\, e^{-i\omega|\tau|} = \frac {2}{i\omega}
\eqno(53)$$
and the $\omega$-integral can be evaluated using the formula

$$\int_{0}^{\infty}\frac {\sin ax}{x}dx = \frac {\pi}{2}, \quad {\rm for}
\quad a>0.
\eqno(54)$$

After using eqs. (53) and (54), we get

$$V({\bf x} - {\bf x'}) =\frac {1}{c} \frac {1}{4\pi}
\frac {1}{|{\bf x} - {\bf x}'|}.
\eqno(55)$$

In case of the massive photon, the mathematical determination of potential
is the analogical to the massless situation only with the difference we use
the propagator (51) and the table integral:

$$\int_{0}^{\infty}\frac {dx}{x}\sin\left(p\sqrt{x^{2}-u^{2}}\right) =
\frac {\pi}{2}e^{-pu}.
\eqno(56)$$

Using this integral we get that the potential generated by the massive
photons is

$$V({\bf x} - {\bf x'},m^{2}) = \frac {1}{c}\frac {1}{4\pi}
\frac {\exp{\left\{-\frac {mcn}{\hbar}|{\bf x} - {\bf x'}|\right\}}}
{|{\bf x} - {\bf x}'|}.
\eqno(57)$$

So we have seen that the basic fact in quantum field theory
was the Green function, then potential and then
force derived from the potential.
The question arises what is the potential between two charges immersed
into the sea of photons of the black body. This problem was solved
(Pardy, 1994g).

\section{The scientific activity}
\hspace{3ex}

The fundamental mechanism of the scientific research was defined by
Gilman (1985):

\begin{quote}
{\bf  "Research work is inherently sequential. It involves overcoming a
series of obstacles that must be approached one after one. This
cannot be avoided because a subsequent obstacle does not become
well-defined until the preceding one has been overcome. Thus, unlike
manufacturing, research work does not lend itself to being accelerated 
by doing several operations in parallel. And the rate at which it
achieves progress depends on the instantaneous effort that is put
into it"}.
\end{quote}

Similarly to the Gilman philosophy expresses the physical credo
David Stern (Stern, 1993).
According to him, scientific work requires continuously to keep notes
of scientific ideas, collect them, transcribe them, illustrate them,
to use word processors in order to edit and produce them.

In order to realize the progress in science it is necessary to
select the right text for study. {\bf A poor text is frustrating, only a
good one makes soar.} It is necessary to select text which provides
intuitive insights. And the best way how to obtain insight is to
solve problems which follows directly from the text or are
invented independently on it.
{\bf The problems must be physically meaningful.} Only
such problems are interesting. It is necessary to solve big problems
because no one care about publishable petty results. There is no need
to have fear from the big problems because every problem can be
divided or splitted  into a sum of small problems easily solvable.
However, what is big and good problem? It obviously depends on the
scientific educations which enables to formulate the big problems from
the big theoretical experience. It depend also on the mentality
of a person who estimates the problem and also on the situation in
the scientific world. {\bf The best way is to have not only one problem
but to formulate many problems. Then, to establish the hierarchy of
them in order to decide what problem is solvable immediately and what
problem is more difficult for solution.}

There is no need to have a fear from drudgery because no pain, no
gain. On the other hand a {\bf minute thinking is more effective than
hours of memorising}.

The effective way is discussion on the similar problems with experts.
At the same time the effective way is to write a review of what was
yet done in the certain field and it forces us to think not only
analytically but also synthetically. It is obvious that {\bf it is
possible to unify all heuristical activities in the maximalistic
form of the scientific publication.} Namely, to write article with the
goal to give clear new mathematical formulation of the new or old
interesting problem, to find solution and to show that the result is
principially experimentally possible to verify. Only in this
situation one has the very strong motivation for dealing with science.
Such approach leads to discoveries in case we use the correct
heuristics. However, the integral part of every heuristics is knowledge of
the process of problem solving. Let us look on this problem.

\section{Problem solving}

We know from the history of mathematics
that Newton approach to the problem solving was expressed by the words:

\begin{quote}
{\bf  "I continuosly keep in the mind the subject of my investigation
  and with patience I wait, unless the first flash little by little
  turns out into the dazzling light"}.
\end{quote}

The deeply true rule for the very effective problem solving and
making the mathematical discoveries was expressed by Poincar\'{e}
(1913) in the following form:

\begin{quote}
{\bf  "The mathematical discovery, the great or the
little one, never arises spontaneously. They ever suppose a ground
into which are peaces of knowledge sewn and very well cultivated by
work, conscious or unconscious"}.
\end{quote}

The substance of the problem solving is also the right question because if we
answer the incorrect question, we move in the blind street.
Solution is the long part of human activity.
Solution of problems is a sequence of orderly steps. However, the
human cognitive actions that arrive at the solutions do not form at
the first moments of such sequence. The sequence is the final result of
the human cognitive process of the problem solving. {\bf The most of the
elementary actions in the problem solving  which do not form the
final chain may consists of false starts, daydreams, idiosyncratic
associations or images}. Only person well trained in the solving
of the particular problems has the experience for the construction of
the final chain from which the start is going to the correct result.

The solution of the problem comes after proper analysis of the
situation and after application of the method "trial and error". This
method is in inevitable integral part of the problem solving.
Nevertheless, solutions are generally {\bf sudden integration}. Usually the
significants of the problem situation come together and are
integrated in a unique way in one pulse of rapid attentional
integration. This characteristic suddenness is usually called as
{\bf insight}. But for this to happen (Blumenthal, 1977),

\begin{quote}
{\bf ``attention must be free enough to go
well beyond the raw data from view due to overloading of the
cognitive capacities"}.
\end{quote}

If we speak about the free attention we mean practically
automatical formation of
of sequences during the problem solving. Automatization of the mental
processes is the fundamental basis for the growth of the development
of the human knowledge. With the automatisation of the cognitive
processes, which means operations without detailed attentional
focusing, attention is free to concentrate on the goals rather than
on the means of performance. The original system of attentional
activity then become automatic tools in the construction of the more
complex mental formation. Or, (Blumenthal, 1977),

\begin{quote}
{\bf  "Our ability to learn may be thus
limited to the degree that our attention is free and to the degree
that it can be directed to the events to be learned"}.
\end{quote}

While on the other hand the well-developed automaticity may at times
block creativity or blind the person to solving the problem, on the
other hand without automaticity there is no progress. The creative
person must be  able to de-automatize and also to apply old automatism in
the new context. According to Whitehead (1928)

\begin{quote}
{\bf  "Civilization advances
to the degree that we can perform important processes without
thinking about them"}.
\end{quote}

The last statement is the idea of the computer creativity.
We work with programs of the
{\bf the computer abstraction, conversation,
proofs of theorems, compilation of texts} and so
on. At present time there are computer chess players and computer
players of the different kinds of plays. We have intelligent programs
such as MATHEMATICA, MAPLE, MATLAB and other superprograms which are
able to solve partial problems of mathematics.
Every computer program is only well organized system of operations in
the microprocessor machine. Of course, we believe that in future
the organization of operations will be so brilliant that computer will
be able solve not only all problems from QED, QCD, gravity, cosmology,
topology and so on,  but also, with some additional
communication with scientists, to create all physical theories and
write the scientific articles and monographs of high quality.

There is no problem for computer to define the imaginary number $i$ as the
solution o the elementary equation $x^{2} + 1 = 0$, and there is no problem
for computer to prove the famous Euler formula

$$i^{i} = e^{-\frac{\pi}{2}}\eqno(58)$$
and other famous Euler formulas.
There is no problem for computer to prove $\sqrt{2} \neq p/q$ where $p$ and $q$
are the integers. On the other hand, to prove that $\sqrt{2} = P/Q$ for the
infinite integers  $P$ and $Q$ is for computer impossible.  The problem
is infinity. Infinity and the infinitesimality is the deadline of
every computer.

\section{General conditions for creativity}

Now, let us investigate the necessary general factors from which the
creative process involving problem solving is built up.
The complete act of the creative
production consists of the four stages: {\bf  Preparation,  Incubation,
Illumination, Verification.}

Other schemes of the creative work was elaborated by Rossman
(Guilford, 1963) who, after study of the 700 productive inventors
concluded that the substantial stages consist in the seven steps.
{\bf  Observation and need of solution of difficulty,
analysis of the need,
survey of all available information, formulation of objective solutions,
critical analysis of the solutions, the birth of the new innovation -
the idea proper, experimentation to test the idea.}

John Dewey gives the following rules for the problem
solving (Guilford, 1963).
{\bf  Recognition of a problem, analysis of a problem, suggestion of
possible solutions, testing of the consequences, judgment of the
selected solution.} So, the similarity between the schemes of
different authors are obvious.

\section{Theoretical physics and technology}

What is the relation between technology
and theoretical physics? Let us remember quotation of
Francis Bacon from his Novum Organon (Rescher, 1978):

\begin{quote}
{\bf  "Neither the naked hand nor the intellect left to itself can effect much.
It is by instruments and helps that work is accomplished, which are as
much needed by the intellect as by hand. And as the instruments of the hand
either impel or guide its motion, so the instruments of the mind
either encourage or admonish the intellect."}
\end{quote}

So, technology forms the inspiration necessary for the theoretical
thinking. Pure technology leads to empirism and alchemy and
pure theory leads to useless symbolic meditation.

Technology is very much older than science. Unaided by science, technology
gave rise to the crafts of primitive man, such as agriculture and
metalworking, the Chinese triumphs of engineering, renaissance cathedrals,
and even the steam engine. Not until the nineteenth century did science
have an impact on technology. In human evolution the ability to make tools,
and so control the environment, was a great advantage, but the ability to do
science was almost entirely irrelevant.

The history of science teaches us that {\bf there exist an escalation of price
of the innovation of science}. But why? Why should the resource-cost
of significant scientific discoveries increase so substantially with the
progress of science? Is this trend only the short-term phenomena
or the structural facts deep-rooted in the nature of science? {\bf What is the
explanation of the cost-escalation and the logarithmic retardation
of the production of scientific results in the modern science? 
What is the explanation of the cost-escalation of LHC?}
Planck's principle is hypothesis that the resource-input costs are
proportional to the scientific innovation. The reasons for the deceleration
of science are speculative, nevertheless there are facts for such
development.

A certain amount of mathematical detail is unavoidable in testing the
model which explains the relationship between science and technology.
It is necessary to have some explicit deployment of certain basic assumptions
about the nature of scientific work. Unless they are set out
with reasonable precision, there is no way of testing the compatibility
and consonance of these assumptions.

Technology is obviously the applied science, and the
technology is crucially dependent upon previous technology (Recher,
1978). For instance the
experimental physics heavily depends on some very advanced
technology. Currently, in fields such as plasma physics and
thermonuclear research, there are many theoretical physicists,
with a broad range of interest and expertise, some with a background
in elementary particle physics, intimately concerned with engineering
questions.

It is non exaggeration to say that the laboratory science
is the result  of the industrial revolution, or the fruit of
its technological harvest.

Just as scientific knowledge and methods are entering technology,
scientific research is in turn determined by technology. Technology
provides instruments and apparatus for scientific investigation
and technological development throws up new fundamental problems that
stimulate the course of scientific research. For instance the transuranic
elements, many isotopes and elementary particles, and most organic
compounds are only created at all through the use of technical aids.

Natural science is fundamentally
empirical and its advance is critically dependent not on human ingenuity
alone, but on the phenomena to which we can only
acces through interactions with
nature. The dependence of the theoretical advance on the goading
of the experimental results was brilliantly expressed by
Max Planck (Rescher, 1978):

\begin{quote}
{\bf  `` ...it was the facts learned from experiments that
shook and finally overthrew the classical theory. Each new idea and each
new step were suggested to investigators -- where it was not
actually thrust upon them -- as the result of measurement}".
\end{quote}

The fundamental research in physics is crucially dependent on the progress
of technology. Historical examples are overhelmingly numerous.
Superconducting magnets for a giant bubble chamber are available
only because of the extremal industrial effort that followed
the discovery of hard superconductors. In experimental physics,
high-energy physics and astronomy  -- wherever photons are counted,
which includes much of fundamental physics -- photomultiplier
technology has often paced experimental progress.
The multidirectional impact of semiconductor technology on
experimental physics is obvious. In several branches of fundamental
physics it extends from the particle detector through
nanosecond circuitry to the computer output of analyzed data.

The creative scientist must be sensitive for
specific situations and he must be able to put the crucial questions.
However, on the other hand he must have a well-developed sense
of the possible as to the issues which, in the existing state of
research technology, one can hope to tackle effectively with the
instruments in hand.

Let us discuss the dependence of the development
of the {\bf particle physics} on the specific technology which enables such
development. The knowledge of the structure of atomic nuclei
and of elementary particles has been derived mainly from the study
of reactions between colliding particles. The method used
is to bombard a "target" with a stream of particles and study
the particles resulting from this interaction. The simplest type
of the reaction is the elastic scattering in which colliding particles simply
deflect each other without undergoing any other change. In this
case the most close distance of approach which can be explored
depends on the initial energy. In inelastic reactions, on the other hand,
one or both of the colliding particles may undergo changes and in addition
new particles and quanta of the electromagnetic radiation may
be created. The creation of new particles and quanta can occur only
if the necessary energy is available in the kinetic energy of the
colliding particles. Consequently, the kinetic energy is the determining
factor for the phenomena to be observed. The high energy particles
occur in nature or as a result of interaction of particles accelerated
in accelerator. Such accelerators have already made available particles
with energy exceeding $2.5 \times 10^{10}$ eV. Particles in a higher
range of energies are accessible only through experiments with
LHC and cosmic rays.

This dependence of {\bf high-energy physics} on technology and
engineering frequently stretches the capabilities of existing
technology to the utmost, requiring innovations and
extrapolations that go well beyond any present state of the art.
Because the resulting technological developments
have implications much broader than their use in the particle physics,
all technology benefits from the opportunity to respond to this
pressure. New technological developments occur sooner
than they would in the absence of such pressure, and they often present
new engineering opportunities, that can be exploited immediately.
Let us recall some examples: very high vacuum systems, sources
of enormous radio frequency power, cryogenic systems,
large-scale static superconducting magnets, pattern recognition devices,
very fast electronic circuits, computers and so on. In this sense,
elementary particle physics has had a major impact on technology,
but the effect was not the result of the direct research.

The development of {\bf special relativity} was evidently conditioned by the
technological success of Michelson and Morley who by their famous
interferometer proved the nonexistence of the aether wind. Without such
technological success, the development of the special relativity was
improbable.

All other experiments which confirmed the relevance of the special relativity
theory, such as the dilation of time, the famous Einstein relation between
mass and energy, the mass dependence on velocity and so on, was evidently
enabled by the high technology and its using in the experimental physics.

Similarly in the general relativity and gravitation the confirmation of the
gravitational waves was conditioned by the gigantic radiotelescope which is
also the success of high technology.
The experimental proof was given by Taylor
and Hulse at the Arecibo radiotelescope.
They performed the systematic measurement of the motion of
the binary with the pulsar PSR 1913+16 (Damour et al. 1992;
Taylor et al. 1992; Taylor, 1993).
They found that the generalized energy-loss formula, which follows
from the Einstein general theory of relativity, is in accordance with
their measurement.
This success was conditioned by the fact that the binary emits
sufficient gravitational radiation to influence the orbital motion of
it at the observable scale.
PSR 1913+16 is now considered as the best general relativistic laboratory .

It is evident that without symbiosis of extremal  high technological
equipment which participated in measurement it was not possible to find that
all data obtained by measurement are compatible with the Einstein quadrupole
formula.

\section{New phenomena and technology}

The really major advances in natural science involve the opening
up of a new point of view or in other words the new way of posing
old problems.

One of the keygoverning conceptions of the contemporary work in
science is the idea of "new phenomena" which falls upon the scientific
community of the day like a bolt from the blue. This idea is handily
developed in a passage by the eminent Russian physicist Piotr Kapitza
(Kapitza, 1964).

\begin{quote}
{\bf  ``I would like to define a "new phenomenon" as a natural phenomenon
that can neither be foreseen nor explained on the basis of existing
theoretical concepts. To clarify the definition, I will name those new
phenomena which, in my opinion, were discovered in the past 150 years.
The first discovery I would like mention is the
discovery of the Galvani electric current in 1789. This discovery in
no way flowed
from the theoretical conceptions of that period. (Volta only half
finished this discovery, if you like). Another discovery worthy of
note, in my opinion, was Oersted discovery in 1820 of the influence
of electric current on a magnetic needle. I do not consider
the Faraday discovery of electromagnetic induction as a new one,
since it is nothing more, than the converse discovery to that
made by Oersted, which it was already possible to foresee.
The Oersted discovery led to Maxwell's equations, and all
others were but an elaboration of this discovery. To foresee the Oersted
discovery theoretically was impossible.

A further example of a new phenomenon is the Hertz discovery of external
photo-effect (not electromagnetic waves which we all know) which
thirty or forty years later led to Einstein's equations, which were
impossible to foresee theoretically. The principles of
indeterminacy and the quantum
theory were contained in the discovery of the photo-effect; all the others
were just a further elaboration.

Then, I would include the Becquerel discovery of radioactivity. This also
was an unforeseable phenomenon. On the other hand,
the discovery of the electron cannot
be counted as an independent discovery. Next, I would note the Michelson
and Morley experiment, the result of which was impossible to foresee.
Then Hertau's discovery of cosmic rays, which was also unforeseeable.
It also appears to me that the discovery of uranium fission by
Lise Maitner and Otto Hahn should be noted}".
\end{quote}

The new data can be derived only
by the new technology and once this technology becomes available there
results a land-office scramble for its exploitation. Nothing more
clearly manifests the dependency of scientific progress upon
advances in the technology of data-aquisition and coordination which create
the conditions requisite for scientific innovations than the common
phenomenon of a clustering of redundant findings. {\bf 
Science grows rapidly around
laboratories, around discoveries which make the testing the hypotheses
easier, which provide the sharp and consistent selective systems.}
Thus the barometer, microscope, telescope, galvanometer, cloud chamber
chromatograph all have stimulated rapid scientific growth.

Nothing could more emphatically demonstrate the nonability of the mere
intellect unaided by the technological means for the acquisition of
empirical data than the fact that nowadays in many areas of natural science
it is virtually impossible that major discoveries or any original work
or real value and interest should come from some quarter outside the
handful of major research groups or institutes that are "on top of the problem"
at hand and are privy to the new data generated by the frontier technology
of research that represents a special "in-house" information source
such as particle accelerator, research reactors, radiotelescope and so on.
The case of Einstein who was able to do discoveries without laboratories
is necessary to discuss apart.

On the other hand the progress without the new data is, of course, possible in
various fields of scholarship and research. The example of pure mathematics
shows that the discoveries can be made in the area of inquiry
that has no empirical datum content at all. But this is not the case
which concerns directly the natural science. Only natural science
depends directly on the empirical data and it differs from the so called
pure sciences as mathematics, logic, or hermeneutics, where in case of
hermeneutics the substance of the intellectual activity consists
in the imaginative reinterpretation  of the old data from the new
conceptual perspectives. Without the influx of new empirical data all
one could do is to proceed to an increasingly sophisticated
conceptual resystematization and reinterpretation of the same
data base. While this might indeed constitute a progress of sorts,
it just is not the sort of progress at issue in natural science,
and blocks the way to the empirical process of testing
the hypothesis and elimination the indirect ideas on which the progress
of science as it is well known standardly depends. The sort of
reinterpretative innovation can only have major significance in the
human sciences where the theleogical concept of meaning is operative.
The reinterpretation might count as "progress" but this sort of
progress will not represent a major advance in the area of
natural science where the strictly hermeneutic issues of meaning
and theleology at best play a minor and peripheral part. Nevertheless
the hermeneutics has the pedagogical values and can be a good
gymnastics for the human brain.

\section{Reciprocity}

The scientific progress in technology depends
undoubtedly on the progress of science and it may be easy to see
that the progress  in science depends on progress in technology.
The relation between science and technology is in such a way
reciprocal (Toulmin, 1972).

\begin{quote}
{\bf `` A natural science and an associated technology are
partners in a kind of historical gavotte, developing most
effectively when their changes are harmoniously synchronized''} .
\end{quote}

The expressive explanation of the relationship between sience
and technology is presented by Hermann Bondi (1967).
He writes:

\begin{quote}
{\bf ``I have spoken earlier
of disproof as the essential agent of progress, but why can we disprove
today what we couldn't disprove yesterday? The answer is that today
we can carry out more accurate experiments relating to matters
that were inaccessible yesterday. We can do so because of the progress
of technology. And so a progressing technology is an absolutely
essential condition for a progressing science. It is a peculiar
disease of this country, I think to feel that science sort of marches
in front and that poor, dirty technology follows a long way behind. But the
relation of science and technology is the relation of the chicken and
the egg; you cannot have the one without the other. It is true that
modern technology derives from modern science, but we would not have
any of modern science without modern technology''.} .
\end{quote}

The historical development of science thus moves in a dialectic
feedback of interchange from the one side to the other of the
theoretical-technological divide, doing so at increasing levels of
sophistication. It can be illustrated by construction of {\bf ITER}
where the compression of matter with laser beams will be probably
used in order to get nuclear fusion (Nakai and Mima, 2004).

ITER means ``the way'' in Latin and it can be transaled as {\bf
 International Thermonuclear Experimental Reactor}. While the problem
 of interaction of an electron with two laser beams was solved only
 theoretically with Dirac equation in the restricted form
(Pardy, 2005c), it leads to the practical use if we will consider more
realistic situation with the laser pulses (Pardy, 2003a).
The practical meaning is, that two laser beams (pulses) impinging
on a target which is constituted from material in
the form of a foam, can replace 100-200 laser beams impinging on a normal
target and it means that the nuclear fusion with two laser beams is
realistic in combination with the thermonuclear reactor
to generate the physical process of implosion (Rozanov, 2004).

Nuclear fusion involves the bringing together of atomic nuclei.
The sum of the individual masses of the nucleons is greater than the
mass of the whole nucleus. This is because the strong nuclear force
holds the nucleons together. Then, the combined nucleus is a lower
energy state than the nucleons separately. The energy difference is
released in the fusion process.

There are two major fusion processes. The magnetic confinement and
inertial confinement. The inertial fusion occurs inside targed fuel
pellets by imploding them with laser or particle beam irradiation in
brief pulses. It produces extremely high densities in the targed where
the laser pulse creates a schock wave in the pellet that it
intensified by its internal geometry. On the other hand, magnetic
fusion devices, like the tokamak, operate at lower densities, but use
magnetic fields to confine the plasma for longer time.

To achieve a burning plasma, a sufficiently high density of
fuel must be heated to temperatures about 100 million degrees of
Celsius that the nuclei collide often enough despite their natural
repulsive forces and energy losses.

The fuels  to be used are two isotopes of hydrogen. Namely, deuterium and
tritium. While deuterium occurs naturally in sea water and it means it
is inexhaustible, tritium can be
bred in a fusion system when the light element, lithium, absorbs
neutrons produced in the fusion reaction. World resources of lithium
are inexhaustible and it means that also the energy obtained by fusion
process is practically infinite.
There are only a few fusion reaction channels in the D-T mixtures in
ITER:

$$ D^{2} + T^{3} \quad \rightarrow \quad He^{4}(3.5 MeV) + n(14.1 MeV)
\eqno(59)$$

$$ D^{2} + D^{2} \quad \rightarrow \quad He^{3}(0.8 MeV) + n(2.5 MeV)
\eqno(60)$$

$$ D^{2} + D^{3} \quad \rightarrow \quad T^{3}(1.0 MeV) + p(3.0 MeV)
\eqno(61)$$

$$ D^{2} + He^{3} \quad \rightarrow \quad He^{4}(3.7 MeV) + p(14.7
MeV).
\eqno(62)$$

Fusion reaction only occur at a sufficient rate for useful power
production when the nucleai have energies in the range 10 - 100 keV and
can collide almost continuosly.

The high temperature fusion is not the final word of the nuclear
physics. The so called {\bf cold fusion} is also considered as a
source of the cheep energy and the development of this strategy is
very perspective using the so called
{\bf sonoluminescence}. Sonoluminescence is
generated by ultrasound which produces microscopical bubbles in a
liquid. The temperature inside each bubble is tens thousands of Kelvin
and it makes possible for some ingredients to generate the fusion.

The very modern approach to future energetics is to get the pure
energy from the reaction of matter and antimatter. Antimatter can be
created during some experiments in particle accelerators and then
included in the special traps. Matter and antimatter forms the
interaction called {\bf annihilation} with the maximal product of
energy. {\bf The so called  annihilation reactor is the reactor without
jeopardy.}

All new approach to energetics laser fusion in ITER, cold fusion,
annihilation reactor represent the reciprocity between theory and
technology. It is evident that the {\bf byproducts of such reciprocity are
many fundamental physical discoveries} generated only by this way.

The reciprocity can be realized also by so called
{\bf theoretical inventions}, which are not still realized but
they are described and suggested
theoretically. Such invention is for instance the so called magnetronic laser
{\bf MAL} for the therapy of cancer. It is proposed in order to show
the alternative way of the therapy of cancer (Pardy, 2003c).

The other possibility of the reciprocity between technology and theory is the
so called laser acceleration of elementary particles. This is very old
problem and QED approach was suggested (Pardy, 1997b, 2001b).

\section{The technological limit}

Progress in science is limited at any given stage of scientific history by the implicit barriers set
by the available technology of data acquisition and processing. Technological
dependency sets technological limits.

The crucial example is the situation in the ancient Greek. From the given
information technology of the day it is not just improbable but actually
inconceivable that the Greek astronomers should have come up with an
explanation for the red shift or the Greek physicians with an account of the
bacteriological transmission of some communicable disease. The reason for
this is simple because the relevant types of data needed to put such
phenomena within the cognitive grasp of man simply lay beyond their reach.
Progress in theorizing in these directions was barred by a technological
barrier on the side of data that was as absolute as the then-extant
technological barriers in the way of developing the internal combustion
engine or the wireless telegraph. So we endorsed the idea that the technology
is crucially conditioned by the deliminative role of the
technological possibilities of the time.

On the other hand {\bf modern technology crucially depends on
the theoretical basis of science}. The obvious  example is
the knowledge of the infinitesimal
calculus. In time of Galileo and Descartes it was a custom to study
the acceleration of a body and its dependence on the distance or time.
Nevertheless, it was absolutely impossible to
develop dynamics of the more complicated motions of bodies because the
mathematical instruments of Galileo or Huygens was not sufficiently
elaborated for the description of the complexity of such motion.
So after invention of
the infinitesimal calculus it was possible to consider
new problems in mechanics
but also in optics in hydrodynamics and so on and to generate the
technological instruments for verification of the behavior of the mechanical,
optical, hydrodynamical and so on systems. Without such subtle instrument as
the infinitesimal calculus was, it was practically no motivation and no
inspiration to invent new equipments.

In such a way, the effective development of science consists in {\bf
the symbiosis of the mathematical instruments with the technology}.
The symbiosis is so strong that practically there is no possibility
for development one of them alone without development of other (Pardy,
1996b).

\vspace{10mm}
\noindent
{\bf References}
\vspace{5mm}

\noindent
Alvarez, L. W. (1973). Certification of three old cosmic-ray emulsion
events as $\Omega^{-}$ decays and interactions, {\it Phys. Rev. D }{\bf 8},
702.\\[2mm]
Bia{\l}ynicky-Birula, I. and  Mycielski, J. (1976).  Nonlinear wave mechanics
{\it Ann. Phys.} ({\it N.Y.}) {\bf 100}, 62.\\[2mm]
Berestetzkii, V. B., Lifshitz,  E. M.  and  Pitaevskii, L. P.
{\it Quantum electrodynamics}, Moscow, Nauka, (in Russian), 1989. \\[2mm]
Blokhintzev, D. I. {\it The fundamental questions of quantum
mechanics}, Academia, Prague, (in Czech), 1971. \\[2mm]
Blumenthal, A. L., {\it The process of cognition}, Prentice Hall,
Inc., Englewood Cliffs, New Jersey, 1977.\\[2mm]
Bodnarczuk, M. (1990). {\it Assuring both quality and creativity in basic
research}, presented at the 17th Annual Conference of the American
Society for Quality Control-Energy Division, Tuscon, Arizona,
September 9-12.\\[2mm]
Bohm, D. and  Vigier, J. (1954). Model of the causal interpretation of
quantum theory in terms of a fluid with irregular fluctuations,
{\it Phys. Rev.} {\bf 96}, 208.\\[2mm]
Bondi, H. Assumption and myth in physical theory,
Cambridge, 1967.\\[2mm]
Damour, T. and  Taylor J. H.  (1992). Strong-field test of relativistic
gravity and binary pulsars, {\it Phys. Rev. D }{\bf 45}, 1868.\\[2mm]
Dirac, P. A. M. (1928a). The quantum theory of electron I,
{\it Proc. Roy. Soc.} {\bf A 117}, 610.\\[2mm]
Dirac, P. A. M. (1928b). The quantum theory of electron II,
{\it Proc. Roy. Soc.} {\bf A 118}, 351.\\[2mm]
Dirac, P. A. M.  (1931). Quantized singularities in the electromagnetic
field , {\it Proc. Roy. Soc.} {\bf A 133}, 60.\\[2mm]
Dirac, P. A. M.  (1939). The relation between mathematics and physics,
{\it  Proc. Roy. Soc., Edinburgh} {\bf 59}, 122.\\[2mm]
Dirac, P. A. M. (1972a). A positive - energy relativistic wave equation I,
{\it Proc. Roy. Soc.} {\bf A 322}, 435.\\[2mm]
Dirac, P. A. M. (1972b). A positive - energy relativistic wave
equation II, {\it Proc. Roy. Soc.} {\bf A 328}, 1.\\[2mm]
Dirac, P. A. M.  (1977a). The relativistic electron wave equation,
Report KFKI - 1977-62, Hung. Ac. Sci. \\[2mm]
Dirac, P. A. M. in: {\it History of twentieth century physics}, 109,
Academic Press, New York, 1977b.\\[2mm]
Dirac, P. A. M. (1982). Pretty mathematics, {\it Int. Journal of
  Theor. Phys.} {\bf 21}, 603.\\[2mm]
Feynman, R. P.  (1963). Quantum theory of gravitation,
{\it Acta Phys. Polonica} {\bf 24}, 697.\\[2mm]
Fok, V. {\it Theory of space, time and gravity}, GIFML, Moscow, 1961.\\[2mm]
Friedman, Y. and Gofman, Y. (2005). Kinematic relations between
relativistically accelerated systems in flat space-time based on
symmetry, e-print gr-qc/0509004.\\[2mm]
Gilman, J. J.  (1985). {\it Important of stable effort for research
productivity}, preprint LBL-20353, September.\\[2mm]
Gupta, S. (1952). Quantization of Einstein gravitational fields: general
treatment, {\it Proc. Phys. Soc.} {\bf A 65}, 608.\\[2mm]
Guilford, J. P. {\it Intellectual factors in productive
thinking}, The second conference on productive thinking,May 2-4,
Washington, D.C., 1963.\\[2mm]
Hadamard, J.  Newton end the infinitesimal calculus, in: The
Royal Society Newton Tercentenary Celebrations. (1946).\\[2mm]
Hawking, S. {\it A brief history of time}, Bantam press, London, 1989.\\[2mm]
Kamran, M. in: Dirac - the taciturn genius, preprint IC/89/265,
Miramare-Trieste, 1989. \\[2mm]
Madelung, E. (1926). Quantentheorie in hydrodynamischer Form,
{\it Z. Physik} 40, 322.\\[2mm]
Kapitza,  P. The future problems of science, in:
{\it The science of science}, ed. by Goldsmith, M. and Mackay, A. (eds.),
London, 1964.\\[2mm]
Kuhn, T. (1963). {\it Interview with Dirac}, May 7, Niels Bohr
Library, Am. Inst. of Phys., New York. \\[2mm]
Lemos,  N. A.  (1983). Stochastic description of the Birula-Mycielski nonlinear equation, {\it Phys. Lett.} {\bf 94A}, (1), 20.\\[2mm]
Nakai, S. and Mima, K. (2004). Laser driven inertial fusion energy:
present and perspective, {\it Rep. Prog. Phys.} {\bf 67}, 321.\\[2mm]
Pais, A. {\it Playing with equation, the Dirac way}, Report
Number DOE/ER/40033B--106, RU/86/150, Rockefeller University, New
York, 1986.\\[2mm]
Pardy, M. (1969). A remark on the clock paradox, {\it Phys. Lett.} {\bf 28 A },
No. 11, 766. \\[2mm]
Pardy,  M. (1973a).
Classical motions of spin 1/2 particles with zero anomalous
magnetic moment, {\it Acta Phys. Slovaca} {\bf 23}, No. 1, 5. \\[2mm]
Pardy,  M. (1973b). WKB approximation for the Duffin-Kemmer equation for spin
zero particles, {\it Acta Phys. Slovaca} {\bf 23}, No. 1, 13. \\[2mm]
Pardy,  M. (1973c). The generalized Feynman integral, {\it Int. Journal of
Theor. Phys.}, Vol. {\bf 8}, No. 1 , 31. \\[2mm]
Pardy,  M. (1983a). Particle production by the \v{C}erenkov mechanism,
{\it Phys. Lett.} {\bf 94A}, No. 1, 30. \\[2mm]
Pardy  M. (1983b). The synchrotron production of gravitons
by the binary system,
{\it General Relativity and Gravitation} {\bf 15}, No. 11, 1027. \\[2mm]
Pardy, M. (1983c). Particle production by accelerated charges moving in
condensed matter, {\it J. Phys. G, Nucl. Phys.} {\bf 9}, 853.\\[2mm]
Pardy, M. (1984). Particle production by charges in condensed matter,
{\it Int. Journal of Theor. Phys.} {\bf 23}, No. 5, 469.\\[2mm]
Pardy, M. (1985). Radiative corrections to classical particle motion
in a magnetic field, {\it Phys. Rev. D} {\bf 31}, No. 2, 325.\\[2mm]
Pardy, M. (1989a). Finite-temperature \v{C}erenkov radiation,
{\it Phys. Lett.} {\bf A 134}, No. 6, 357.\\[2mm]
Pardy, M. (1989b). The quantum states of the superfluid ring from the
Gross-Pitaevskii equation, {\it Phys. Lett.} {\bf A 140}, Nos. 1, 2, 51.\\[2mm]
Pardy, M. Possible tests of the nonlinear quantum mechanics,
in: {\it Waves and Particles in Light and Matter}, Ed. A. Garuccio and
A. van der Merwe, Plenum Publishing Corporation, 419-422, 1994a.\\[2mm]
Pardy, M. (1994b). The \v{C}erenkov effect with radiative corrections,
Phys. Lett. {\bf B 325}, 517\\[2mm]
Pardy, M.  (1994c). The synchrotron production of photons with
radiative corrections, {\it Phys. Lett.} {\bf A 189}, 227.\\[2mm]
Pardy,  M. (1994d). The high-energy gravitons from the binary, e-print:
CERN-TH.7239/94\\[2mm]
Pardy, M. (1994e). The gravitational \v{C}erenkov radiation with
radiative corrections, {\it Phys. Lett.} {\bf B 336}, 362;
ibid: e-print CERN-TH.7270/94. \\[2mm]
Pardy,  M.  (1994f). The emission of photons by plasma fluctuations, e-print
CERN-TH.7348/94.  \\[2mm]
Pardy, M. (1994g). The two-body potential at finite temperature,
e-print CERN-TH.7397/94. \\[2mm]
Pardy,  M. (1995). The finite-temperature gravitational \v{C}erenkov radiation,
{\it Int. Journal of Theor. Phys.} {\bf 34}, No. 6, 951.\\[2mm]
Pardy,  M. (1996a). The string model of gravity, gr-qc/9602007.\\[2mm]
Pardy,  M. (1996b). Technological support of theoretical sciences,
{\it Scripta Fac. Sci. Nat. Univ. Purk. Brun.} {\bf 24 -- 26},
No.7, 67. \\[2mm]
Pardy,  M. (1997a). \v Cerenkov effect and the Lorentz contraction,
{\it Phys. Rev.} {\bf A 55}, No. 3. 1647.\\[2mm]
Pardy,  M. (1997b). The quantum field theory of laser acceleration,
{\it Phys. Lett.} {\bf A 243}, 223. ibid. hep-ph/0005319. \\[2mm]
Pardy,  M. (2000a). The two-body electromagnetic pulsar, hep-ph/0001277.\\[2mm]
Pardy,  M. (2000b). Synchrotron production of photons by a
two-body system, {\it Int. Journal of Theor. Phys.} {\bf 39}, No. 4, 1109;
hep-ph/0008257.\\[2mm]
Pardy,  M. (2001a). Deflection of light by the screw dislocations in
space-time, gr-qc/0106019.\\[2mm]
Pardy,  M. (2001b). The quantum electrodynamics of laser acceleration,
{\it Radiation Physics and Chemistry}  {\bf 61}, 391.\\[2mm]
Pardy,  M. (2001c). The circle electromagnetic pulsar, hep-ph/0111036. \\[2mm]
Pardy,  M. (2001d). To the nonlinear quantum mechanics,
quant-ph/0111105.\\[2mm]
Pardy,  M. (2002a). \v Cerenkov effect with massive photons, {\it Int.
Journal of Theor. Phys.} {\bf 41}, No. 5, 887.\\[2mm]
Pardy,  M. (2002b). Largest terrestrial electromagnetic pulsar,
{\it Int. Journal of Theor. Phys.} {\bf 41}, No. 6, 1155.  \\[2mm]
Pardy,  M. (2003a). Electron in the ultrashort laser pulse, {\it Int.
Journal of Theor. Phys.} {\bf 42}, No. 1, 99. ibid: hep-ph/0207274.\\[2mm]
Pardy,  M. (2003b). The space-time transformations between accelerated
systems, gr-qc/0302007.\\[2mm]
Pardy,  M. (2003c). Theory of the magnetronic laser, physics/0306024.\\[2mm]
Pardy,  M. (2003d). Massive photons in particle and laser physics,
hep-ph/0308190.\\[2mm]
Pardy,  M. (2004a). The space-time transformations and maximal acceleration,
{\it Spacetime \& Substance Journal} {\bf 1}(21), 17.\\[2mm]
Pardy, M. (2004b). Massive photons and the Volkov solution,
{\it Int. Journal of Theoretical Physics} {\bf 43}, No. 1, 127.\\[2mm]
Pardy, M. (2004c). Volkov solution for an electron in the two wave fields,
hep-ph/0408288.\\[2mm]
Pardy, M. (2004d). Radiation of the gravitational and electromagnetic pulsars,
hep-ph/0412330.\\[2mm]
Pardy,  M. Light in metric space-time and its deflection by the screw
dislocation, {\it Spacetime Physics Research Trends. Horizons in World
Physics}, Volume {\bf 248}, ISBN 1-59454-322-4, 2005a.
ibid. e-print hep-ph/0402249. \\[2mm]
Pardy,  M. (2005b). The propagation of a pulse in the real strings and rods,
math-ph/0503003.\\[2mm]
Pardy, M. (2005c). Radiation of the black body in the
external field, quant-ph/0509081 \\[2mm]
Pietschmann, H. (1978). The rules of scientific discovery
demonstrated from examples of the physics of elementary particles,
{\it Foundation of physics} {\bf 8}, Nos. 11/12, 905.\\[2mm]
Poincar\'{e}, A. (1913). {\it Science and Method},
Thomas Nelson and Sons, London, (1902).\\[2mm]
Rafanelli, K, and Schiller, R. (1964). Classical motion of
spin-1/2 particles, {\it Phys. Rev.} {\bf 135}, No. 1 B, B279.\\[2mm]
Rescher, N. {\it Scientific progress}, Basil Blackwell, Oxford, 1978.\\[2mm]
Ritus, V. I. (1979). {The quantum effects of the interaction of
elementary particles with the intense electromagnetic field},
{\it Trudy FIAN} {\bf 111}, 5.\\[2mm]
Rosen, N. A. (1974). Classical picture of quantum mechanics,
{\it Nuovo Cimento} {\bf 19B},(1) 90.\\[2mm]
Rozanov, V. B. (2004). On the possible realization of spherical
compression for fusion targets irradiated by two laser beams, {\it Uspekhi
Fiz. Nauk}, {\bf 174}(4), 371. (in Russian).\\[2mm]
Salam, A. {\it From a life of physics}, International atomic agency.\\[2mm]
Santilli, R. M. (2005). Inconsistencies of general relativity and
their apparent resolution via the Poincar\'{e} invariant
isogravitation, IBR-TPH-03-05.\\[2mm]
Schwinger, J.  (1976). Gravitons and photons, the methodological unification
of source theory, {\it General Relativity and Gravitation.}
{\bf 7}, No, 3, 251.\\[2mm]
Stern, D. (1993). All I really need to know, {\it Physics Today}, 63.\\[2mm]
Taylor, J. H. Jr. (1993). Binary pulsars and
relativistic gravity (Nobel Lecture).\\[2mm]
Taylor, J. H. Wolszczan, A. Damour T. and  Weisberg J. M. (1992). Experimental
constraints on strong-field relativistic gravity,
{\it Nature} {\bf 355}, 132 .\\[2mm]
Toulmin, S. {\it Human Understanding}, Vol 1., (Princeton), p. 373,
(1972). \\[2mm]
Whitehead, A. N. {\it Process and Reality}, New York, Macmillan. \\[2mm]
Weisskopf, W. (1971). {\it My life as a physicist},
Lecture given at the Erice Summer
School in High-Energy Physics, Sicily, Italy in 1971. \\[2mm]
Wolpert, L. {\it The unnatural nature of science}, Faber and Faber,
London, Boston, 1993.
\end{document}